\documentclass{article}

\usepackage[preprint]{neurips_2026}

\usepackage[utf8]{inputenc}
\usepackage[T1]{fontenc}
\usepackage{hyperref}
\usepackage{url}
\usepackage{booktabs}
\usepackage{graphicx}
\usepackage{amsfonts}
\usepackage{array}
\usepackage{tabularx}
\usepackage{float}
\usepackage{nicefrac}
\usepackage{microtype}
\usepackage{xcolor}
\usepackage{enumitem}
\usepackage{fancyvrb}
\usepackage{mdframed}
\usepackage{listings}
\lstdefinestyle{kisspy}{
  language=Python,
  basicstyle=\ttfamily\scriptsize,
  keywordstyle=\color[rgb]{0.0,0.0,0.8}\bfseries,
  commentstyle=\color[rgb]{0.25,0.5,0.35}\itshape,
  stringstyle=\color[rgb]{0.73,0.13,0.13},
  showstringspaces=false,
  frame=single,
  framesep=2mm,
  numbers=left,
  numberstyle=\tiny\color{gray},
  numbersep=5pt,
  breaklines=true,
  tabsize=4,
  captionpos=b,
}
\lstnewenvironment{prompt}{%
  \lstset{language={},basicstyle=\ttfamily\scriptsize,frame=single,framesep=2mm,breaklines=true,columns=fullflexible}%
}{}
\newenvironment{promptbox}{%
  \begin{mdframed}[
    linewidth=0.4pt,
    linecolor=black,
    innerleftmargin=5pt,
    innerrightmargin=5pt,
    innertopmargin=5pt,
    innerbottommargin=5pt,
    skipabove=2pt,
    skipbelow=5pt]
  \setlength{\parskip}{0pt}\setlength{\parindent}{0pt}%
  \ttfamily\scriptsize\raggedright\frenchspacing\obeylines\leavevmode\strut\ignorespaces
}{%
  \end{mdframed}
}

\title{KISS Sorcar: A Stupidly-Simple General-Purpose and Software Engineering AI Assistant}

\author{%
Koushik Sen \\ EECS Department, UC Berkeley\\ \texttt{ksen@berkeley.edu} \\ }

\begin{document}

\maketitle

\begin{center}
\textit{"Everything should be made as simple as possible, but not simpler."}\\
\textit{Albert Einstein}
\end{center}

\begin{abstract}
Large language models can generate code and call tools fluently, yet deploying them as practical assistants for \emph{long-horizon} software engineering and AI-discovery tasks still exposes persistent gaps: finite context windows, a single mistake that can derail entire sessions, agents that get stuck in dead ends, AI slop, and generated changes that are difficult to review or revert.

We present \textbf{KISS Sorcar}, an \emph{open-source general-purpose AI agent for long-horizon tasks and AI discovery} that doubles as an integrated development environment (IDE).  It is built on top of the \textbf{KISS Agent Framework}, a stupidly-simple AI agent framework of roughly 2,900~lines of code for the core agents.  The framework addresses the gaps above through a structured system prompt and a five-layer agent hierarchy in which each layer adds exactly one concern: budget-tracked ReAct execution, automatic continuation across sub-sessions via summarization, coding and browser tools with parallel sub-agents, persistent multi-turn chat with history recall, and git worktree isolation so every task runs on its own branch.  Engineering principles are encoded in the agent's system prompt.

KISS Sorcar is a free, simple, local-first, bring-your-own-key assistant that ships as three coordinated surfaces over a single local daemon: a Visual Studio Code extension, a Claude-Code-style CLI (\texttt{sorcar}), and a browser/mobile web app; all agents run as daemons.  Prompts and code are sent directly to the model provider or local endpoint the user configures, not through any intermediary servers.  The framework supports mixing models from multiple vendors in the same task simply via prompts (OpenAI, Anthropic, Gemini, Together, Z.AI, Moonshot AI, OpenRouter, Claude Code CLI, and Codex CLI), bundles a catalog of 504~models across 9~provider categories, includes 23~third-party messaging agents (Slack, Gmail, WhatsApp, SMS, phone-call control, and others), discovers Model-Context-Protocol (MCP) servers, loads Agent Skills, and supports browser automation, multimodal input, and Docker containers.

In this research, we deliberately prioritize output quality over speed: giving a frontier model adequate time to validate its own output (running linters, type checkers, and tests) reduces the low-quality code common in faster but less thorough agents.  The entire system was built using itself in 4 months, providing a continuous stress test in which any bug was patched as it appeared.  On Terminal~Bench~2.0, KISS Sorcar achieves a 62.2\% overall pass rate with Claude Opus~4.6, compared with Claude~Code (58\%) and Cursor Composer~2 (61.7\%).  These results are notable because we did not tune our prompts or any model specifically for Terminal Bench~2.0.
\end{abstract}

%----------------------------------------------------------------------
\section{Introduction}
\label{sec:intro}
%----------------------------------------------------------------------

Modern Large language models (LLMs), such as Anthropic's Claude Opus 4.7~\citep{anthropic2026opus47}, OpenAI's GPT 5.5~\citep{openai2026gpt55}, and Google's Gemini 3.1~\citep{google2026gemini31}, can generate code, reason about software architecture, and use developer tools~\citep{chen2021evaluating,roziere2023code}.  A growing body of work has explored how to harness these capabilities for autonomous software engineering, from single-session agents that resolve GitHub issues~\citep{yang2024sweagent,wang2024openhands} to industrial products marketed as AI software and general assistants ~\citep{copilot2021,cursor2024,devin2024,claudecode2025,codex2025,openclaw2025}.  Yet using an LLM as a practical \emph{general-purpose} assistant for long-horizon software engineering and AI-discovery tasks still exposes several stubborn gaps: context windows are finite, a single mistake can derail an entire session, agents get stuck in dead ends, models generate AI slop, and generated changes are difficult to review or revert once applied to a live codebase.

We propose the \emph{KISS Agent Framework}, a stupidly simple AI agent framework containing roughly 2,900 lines of code for the core agent implementation.  We try to address the above-mentioned gaps through a structured system prompt (Section~\ref{sec:system-prompt}) and a five-layer agent hierarchy in which each layer solves exactly one concern:

\begin{enumerate}[nosep]
  \item \textbf{KISS Agent}: budget-tracked
ReAct~\citep{yao2023react} loop with native function calling.
  \item \textbf{Relentless Agent}: automatic summarization and continuation 
across sub-sessions.
  \item \textbf{Sorcar Agent}: coding tools, browser automation,
and parallel sub-agent execution.
  \item \textbf{Chat Sorcar Agent}: persistent multi-turn chat
sessions with history recall.
  \item \textbf{Worktree Sorcar Agent}: git worktree isolation so
every task runs on its own branch.
\end{enumerate}

The name ``KISS'' reflects the \emph{Keep It Simple, Stupid} design philosophy in software engineering: each layer is small, each concern is isolated, and the overall system avoids unnecessary abstraction.

\begin{figure}[t]
\centering
\includegraphics[width=\textwidth]{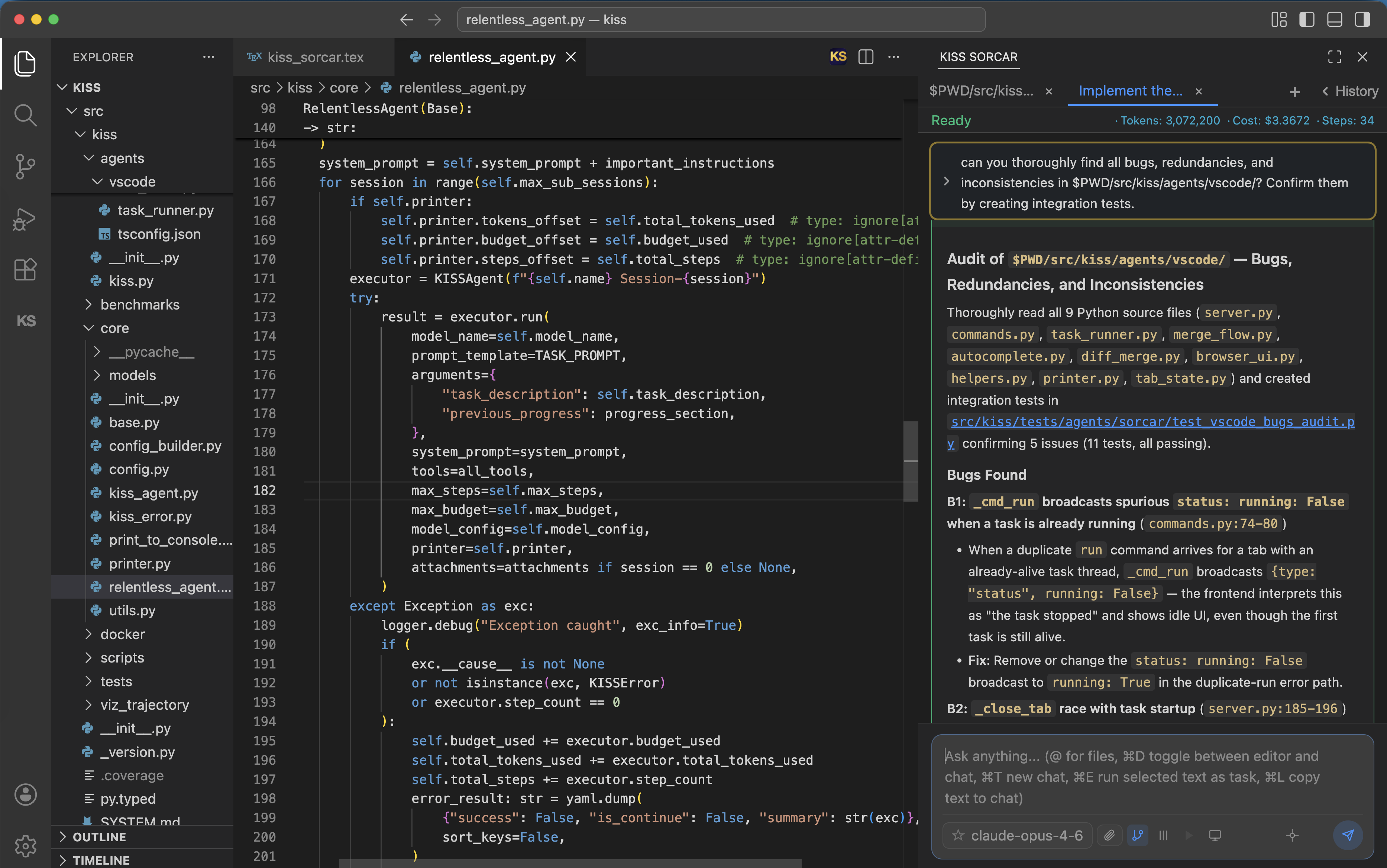}
\caption{Screenshot of KISS Sorcar running as a VS~Code extension. The sidebar shows the agent's chat interface with real-time budget tracking, while the editor displays the code being modified.}
\label{fig:kiss-sorcar-screenshot}
\end{figure}

Table~\ref{tab:assistant-comparison} shows a high-level comparison of KISS Sorcar with Claude Code and Cursor.

\begin{table}[H]
\centering
\caption{KISS Sorcar compared with Claude Code and Cursor, reproduced from the project README.}
\label{tab:assistant-comparison}
\scriptsize
\setlength{\tabcolsep}{3pt}
\renewcommand{\arraystretch}{1.15}
\begin{tabularx}{\textwidth}{>{\raggedright\arraybackslash}p{0.22\textwidth}>{\raggedright\arraybackslash}X>{\raggedright\arraybackslash}X>{\raggedright\arraybackslash}X}
\toprule
\textbf{Capability} & \textbf{KISS Sorcar} & \textbf{Claude Code} & \textbf{Cursor} \\
\midrule
\textbf{Interfaces} & CLI + VS Code extension + web/mobile app & CLI + mobile app & Custom VS Code \\
\textbf{AI Discovery} & \(\checkmark\) simply via prompt & \(\times\) & \(\times\) \\
\textbf{GEPA Prompt Optimization} & \(\checkmark\) simply via prompt & \(\times\) & \(\times\) \\
\textbf{Multiple models from multiple vendors in the same task} & \(\checkmark\) Mix OpenAI, Anthropic, Gemini, Together, Z.AI, Moonshot AI, OpenRouter, Claude Code CLI, and Codex CLI & \(\times\) Anthropic Claude models only & \(\times\) One model per task \\
\textbf{Primary focus} & \(\checkmark\) \textbf{Quality}: rigorous review, end-to-end tests & Speed and developer ergonomics & Speed \\
\textbf{Core Agents \# LoC} & \textbf{\textasciitilde{}2900} & Unknown & Unknown \\
\textbf{Models in bundled catalog} & 504 across 9 provider categories & Claude family only & Subset chosen by Cursor \\
\textbf{Bring your own API key / endpoint} & \(\checkmark\) Yes: keys stay on your machine & \(\checkmark\) Anthropic key & \(\triangle\) Routed through Cursor backend \\
\textbf{Open source} & \(\checkmark\) Apache-2.0 & \(\times\) Proprietary & \(\times\) Proprietary \\
\textbf{Price} & Free framework; pay only your chosen model provider & Subscription / API usage & Subscription \\
\textbf{Run on top of Claude Code / Codex CLI} & \(\checkmark\) \texttt{cc/*} and \texttt{codex/*} namespaces & N/A & \(\times\) \\
\textbf{Messaging and communication channels} & \(\checkmark\) 23 third-party agents, including Slack, Gmail, Phone Control, SMS, and WhatsApp & Partial: Slack, mobile Remote Control, and research-preview channels for Telegram, Discord, and iMessage; no documented built-in Gmail, WhatsApp, phone-call, or SMS channel & Partial: Slack and Microsoft Teams Cloud Agent integrations; no documented built-in Gmail, WhatsApp, phone-call, or SMS channel \\
\textbf{Terminal Bench 2.0 score} & \textbf{62.2\%} & 58\% & 61.7\% (Cursor agent) \\
\bottomrule
\end{tabularx}
\end{table}

We implement KISS Sorcar, an open-source general-purpose AI agent for long-horizon tasks and AI discovery, which doubles as an integrated development environment (IDE) on top of the KISS Agent Framework.  KISS Sorcar is local-first and bring-your-own-key: prompts and code are sent directly to the model provider or local endpoint the user configures, not through any intermediary servers.  It ships as three coordinated surfaces over a single local daemon: a Visual Studio Code extension, a Claude-Code-style command-line interface (\texttt{sorcar}), and a browser/mobile web app, all running locally; all agents run as daemons.  The CLI has both an interactive REPL mode (with slash commands such as \texttt{/help}, \texttt{/clear}, \texttt{/resume}, \texttt{/model}, \texttt{/cost}, \texttt{/skills}, \texttt{/mcp}, \texttt{/autocommit}, and file/folder \texttt{@}-mentions) and a one-shot non-interactive mode (\texttt{-t}/\texttt{-f}).  KISS Sorcar has browser support (using open-source Chromium and Playwright), multimodal support, Docker container support, OpenClaw-like features (whose discussion is beyond the scope of the paper), a mobile/web app, integrated MCP server management (\texttt{sorcar mcp add/list/auth/debug/\ldots}), and a library of Agent Skills loaded from \texttt{\textasciitilde{}/.kiss/skills} and project-local directories.  KISS Sorcar is free and open-source (Apache-2.0), distributed on PyPI as \texttt{kiss-agent-framework}; all one needs is a model API key from a major LLM provider, such as Anthropic, OpenAI, Google, Together, Z.AI, Moonshot AI, or OpenRouter, or a custom local endpoint configured via \texttt{--endpoint}/\texttt{--header}.  The framework can also run on top of the Claude Code CLI (\texttt{cc/*} models) and the Codex CLI (\texttt{codex/*} models), and mix models from multiple vendors within a single task simply by issuing prompts.  We implemented this framework in roughly 4 months, and the repository is available at \url{https://github.com/ksenxx/kiss_ai}. The name ``Sorcar'' pays homage to P.~C.~Sorcar, the Bengali magician.  The engineering principles described in Section~\ref{sec:system-prompt} are encoded in the agent's system prompt.

KISS Sorcar has been built using itself.  The entire codebase (the KISS Agent framework, the Sorcar agent layers, the VS~Code extension, and the system prompt) was developed by KISS Sorcar operating on its own repository.  This self-hosting discipline provides a continuous-integration-style stress test: if the agent introduces a bug that impairs its ability to function, we ask the agent to fix it by analyzing the trajectory and code.  The simplicity of the layered architecture was both a prerequisite for and a consequence of this bootstrapping process. The simplicity of the framework reduced the number of bugs the agent introduced.  The five core agent classes remain compact: the KISS Agent comprises 463~lines, the Relentless Agent 431~lines, the Sorcar Agent 685~lines, the Chat Sorcar Agent 497~lines, and the Worktree Sorcar Agent 848~lines, a total of roughly 2,924~lines of code (counting only significant lines, i.e.\ excluding empty lines, comment-only lines, and docstrings of private methods).  Despite the Sorcar and Chat layers absorbing Docker-aware tool variants, parallel sub-agent orchestration, persistent chat-session bookkeeping, MCP integration, slash-command dispatch, and worktree concurrency safety, the framework has remained within this compact footprint and within the layered, single-concern design.

In the project, we deliberately prioritize output quality over speed.  In our experience, using a weaker or cheaper model often forces the developer to discard the agent's work and retry, ultimately increasing the total cost of completing a task.  Conversely, giving a frontier model adequate time to validate its own output (running linters, type checkers, and tests before declaring success) reduces the ``slop'' (low-quality, subtly incorrect code) common in faster but less thorough agents.  We expect token costs and inference latencies to continue to fall~\citep {gao2025trae}, making this quality-first posture increasingly practical.  In the meantime, the code produced by our agent is well-organized, simple, and idiomatic.

We evaluate on Terminal Bench~2.0 and achieve a 62.2\% overall pass rate using Claude Opus~4.6, compared with Claude~Code (58\%) and Cursor Composer~2 (61.7\%)~\citep{cursor2026composer2} on the same benchmark (Section~\ref{sec:evaluation}).  These results are notable because we did not tune our prompts or any model specifically for the Terminal Bench 2.0.

In KISS Sorcar, we deliberately kept the system simple.  We included only the agent technologies necessary for KISS Sorcar to function as a general-purpose software engineering assistant. We showed that a simple agent framework, without sophisticated agent technologies such as trajectory compaction and asynchronous multi-agent orchestration, was sufficient to build KISS Sorcar.  By building KISS Sorcar using itself and matching or exceeding both Cursor and Claude Code, we found that established software engineering techniques and principles are important for building reliable agent systems.

\paragraph{Outline.}
Section~\ref{sec:architecture} presents the five-layer agent architecture and its motivating design principles. Section~\ref{sec:long-running} describes how the framework supports AI discovery, GEPA prompt optimization, and repository optimization through single-prompt drivers. Section~\ref{sec:evaluation} reports evaluation results on Terminal Bench~2.0.  Section~\ref{sec:vscode-features} covers the user-facing features of the VS~Code extension, CLI, and web app. Section~\ref{sec:system-prompt} details the system prompt. Section~\ref{sec:painless} illustrates painless software engineering through a real development session. Section~\ref{sec:related} discusses related work, and Section~\ref{sec:conclusion} concludes.

%----------------------------------------------------------------------
\section{Agent Architecture}
\label{sec:architecture}
%----------------------------------------------------------------------

We initially built the KISS Agent Framework to rapidly prototype and experiment with various prompt optimization techniques, such as Gepa~\citep{agrawal2025gepa}, and evolutionary algorithms for algorithmic and code optimization, such as AlphaEvolve~\citep{novikov2025alphaevolve} and OpenEvolve~\citep{openevolve2025}. We focused heavily on keeping the agent framework simple so we could rapidly prototype and experiment with ideas.  The simplicity of the framework also enabled coding agents to write simple, bug-free code.  We ultimately did not use any prompt optimization techniques when creating the system prompt for KISS Sorcar, as we hand-tuned it based on our long-term experience with KISS Sorcar and its behavior.  The KISS Agent Framework uses five agent layers in a layered architecture combining composition and inheritance. Each layer delegates upward for the concerns it does not own.

%......................................................................
\subsection{KISS Agent}
\label{sec:kiss-agent}
%......................................................................

The KISS Agent is the innermost execution unit.  It implements a standard ReAct loop~\citep{yao2023react} with the following characteristics.  

\begin{figure}[H]
\begin{lstlisting}[style=kisspy,numbers=none,caption={A complete KISS agent with a single tool.},label={lst:simple-agent}]
from kiss.core.kiss_agent import KISSAgent

def calculate(expression: str) -> str:
    """Evaluate a math expression."""
    return str(eval(expression))

agent = KISSAgent(name="Math Buddy")
result = agent.run(
    model_name="gemini-2.5-flash",
    prompt_template="Calculate: {question}",
    arguments={"question": "What is 15% of 847?"},
    tools=[calculate]
)
print(result)  # 127.05
\end{lstlisting}
\end{figure}

\textbf{Native function calling.}  We register tools as ordinary Python callables.  The agent builds an OpenAI-compatible tool schema once at setup time and caches it, avoiding redundant schema construction on every LLM call.  A special \texttt{finish} tool signals task completion and returns the result to the caller.

\textbf{Step, token, and budget tracking.}  At every step, the agent extracts input and output token counts from the API response, computes the dollar cost using a per-model pricing table, and updates both a local budget counter and a global (cross-agent) budget counter protected by a class-level lock.  The agent checks three limits before each step: the per-agent budget, the global budget, and the maximum step count.

\textbf{Error resilience.}  The agent retries transient API errors (rate limits, server errors) up to a configurable threshold of consecutive failures.  It detects non-retryable errors (authentication failures, permission denials) and raises them immediately.

\textbf{Non-agentic mode.}  When tools are not needed, the agent can run a single generation without the ReAct loop, which is useful for summarization or question-answering sub-tasks.

Listing~\ref{lst:simple-agent} shows a complete, working agent in under ten lines of code.  The developer defines an ordinary Python function (\texttt{calculate}), instantiates a \texttt{KISSAgent}, and calls its \texttt{run} method with a model name, a prompt template, template arguments, and a list of tools.  The framework automatically handles tool-schema generation, the ReAct loop, and budget tracking.

The KISS Agent is stateless across runs: each call to its run method resets the conversation, token counters, and tool registry.  This makes it safe to reuse a single agent instance for multiple sequential tasks.

%......................................................................
\subsection{Relentless Agent}
\label{sec:relentless-agent}
%......................................................................

The Relentless Agent wraps a KISS Agent in a continuation loop.  Its core contribution is the ability to execute tasks that exceed a single context window by breaking them into sub-sessions.

Rather than investing in context-compaction techniques, we adopt a simple continuation protocol: when a sub-session exhausts its context window or step budget, the agent produces a \emph{structured summary} of every action taken so far (chronologically ordered, with explanations and relevant code snippets) and a fresh sub-session resumes from that summary.  This approach is related in spirit to Reflexion~\citep{shinn2023reflexion}, which feeds verbal self-critiques back into subsequent trials; we adapt the idea to continue an unfinished task across sub-sessions rather than to retry from scratch.  While developing KISS Sorcar, we found in our experience that a na\"ive instruction to ``summarize the current context'' produced poor continuations; requiring a \emph{step-by-step chronological account with code snippets} improved coherence across sub-sessions. A potential limitation is that summaries may grow unwieldy for multi-day tasks; in practice, we have not encountered this problem even for tasks spanning several hours, but a thorough evaluation of summary scaling remains future work.

\textbf{Continuation protocol.}  The \texttt{finish} tool exposed to the inner KISS Agent accepts three fields: a success flag, a continue flag, and a summary.  When the agent sets \texttt{is\_continue=True}, the Relentless Agent starts a new sub-session with a fresh context window.  The prompt for the new session includes a chronologically ordered list of all prior attempt summaries and instructs the agent not to redo completed work.  The continuation prompt template is:

\begin{prompt}
# Task Progress (Continuation {continuation_number})

{progress_text}

# Continue
- Complete the rest of the task.
- **DON'T** redo completed work.
- If you have been retrying the same approach without progress,
  step back and rethink the strategy from scratch.
\end{prompt}

\textbf{Forced continuation on failure.}  If a sub-session raises an exception (e.g., the step limit is hit before the agent calls finish), the Relentless Agent does not abort.  Instead, it saves the full trajectory to a temporary file, spawns a separate summarizer agent to read the trajectory and produce a concise summary, and then uses that summary as the progress text for the next sub-session.  This ensures that even crashed sessions contribute useful context to subsequent attempts.  The summarizer receives the following prompt:

\begin{prompt}
# Summarizer

The trajectory of the agent is stored in the file: {trajectory_file}

# Instructions
- Read the trajectory file and analyze it.  The trajectory file
  could be large.
- Return a precise chronologically-ordered list of things the
  agent did with the reason for doing that along with relevant
  code snippets
\end{prompt}

\noindent To force the agent to self-continue before hitting the step limit, we augment the system prompt with an instruction that fires near the end of the budget:

\begin{prompt}
# MOST IMPORTANT INSTRUCTIONS
- **At step {step_threshold}: you MUST call
  finish(success=False, is_continue=True,
  summary="precise chronologically-ordered list of things
  the agent did with the reason for doing that along with
  relevant code snippets")** or if the task is not complete
  and you are at risk of running out of steps or context
  length.
- Work dir: {work_dir}
- Current process PID: {current_pid} -- NEVER kill this
  process.
\end{prompt}

%......................................................................
\subsection{Sorcar Agent}
\label{sec:sorcar-agent}
%......................................................................

The Sorcar Agent adds the tools that make the system useful for software development and general-purpose automation.

\textbf{Coding tools.}  We provide four core tools: a shell command executor with streaming output, a file reader, a precise string-based file editor, and a file writer.  The shell executor supports a configurable timeout, streams output to the user interface in real time, and respects a stop event that allows the user to cancel a running command.  Note that we kept the tool names (Bash, Edit, Write, Read) the same as in Claude Code because the underlying Anthropic models do not make mistakes with these tool names.

\textbf{Browser automation.}  A web-use tool provides programmatic browser control: navigating to URLs, reading page accessibility trees, clicking elements, typing text, pressing keys, scrolling, and taking screenshots.  This enables the agent to research documentation, verify deployed applications, and interact with web-based tools. It uses the open-source Chromium browser using the Playwright library. 

\textbf{Parallel sub-agents.}  A parallel execution tool spawns independent Sorcar Agent instances in a thread pool.  Each sub-agent gets its own LLM context and tool set.  This is useful for embarrassingly parallel tasks such as summarizing multiple files or researching independent topics.  We collect the results and return them in input order.  

\textbf{User interaction.}  An ask-user-question tool allows the agent to pause execution and request clarification from the user.  In the VS~Code integration, this renders as a text input in the sidebar; in CLI mode, it reads from standard input.

\textbf{Docker isolation.}  When a Docker image is specified, we replace the coding tools with Docker-aware variants that execute commands inside a container, providing an additional layer of sandboxing for untrusted tasks.

\textbf{Dynamic model switching with \texttt{set\_model}.}  A \texttt{set\_model} tool, registered automatically on every Sorcar Agent, lets the agent hand the live conversation off to a different LLM at any point during a task without restarting it.  The tool accepts a model name (for example \texttt{"gpt-5.5"}, \texttt{"claude-opus-4-7"}, or \texttt{"gemini-2.5-flash"}), constructs the new backend through the same factory that built the original model, copies the conversation history and the cumulative usage counters into the new backend, rebuilds the cached tools schema in the new provider's dialect (Anthropic and OpenAI schemas differ in subtle ways), persists the choice to disk so that the next task in the same chat session reuses it, and returns a confirmation string back to the agent loop.  If no model has been instantiated yet (deferred-startup case), it simply updates the agent's default \texttt{model\_name} for the next run; if the requested model is already active, it is a no-op.

This single tool is what enables \emph{multi-model, multi-vendor workflows inside a single task}.  Three idiomatic patterns recur in our usage: (i)~a \emph{scout-then-edit} pattern where a cheap fast model (e.g.\ \texttt{gemini-2.5-flash}) reads the repository, grep-searches for relevant call sites, and gathers context, then \texttt{set\_model("claude-opus-4-7")} hands off to a stronger reasoner for the actual edit; (ii)~a \emph{generate-then-review} pattern where a primary model produces a change and the agent calls \texttt{set\_model("gpt-5.5")} to switch to a second-opinion model that re-reads the diff, runs the tests, and verifies the change against the original request; and (iii)~a \emph{cost-aware long-loop} pattern where an open-ended optimization or discovery task (Section~\ref{sec:long-running}) spends a cheap model on bookkeeping and journal-keeping but pays for a frontier model only at the decision points.  Because the conversation transcript and usage counters are carried across the switch, all subsequent budget accounting and continuation summaries remain coherent regardless of how many times the model is changed.

The following three prompts, reproduced from \texttt{src/kiss/INJECTIONS.md}, illustrate what this tool layer empowers.  They are not API calls or special modes: each is a single natural-language instruction that can be pasted into the IDE or CLI.  Their power comes from the combination of dynamic model switching, coding tools, test execution, parallel sub-agents, and later layers' persistence and worktree isolation.

\noindent\textbf{Primary model plus independent review.}\par\vspace{1pt}
\begin{promptbox}%
Use claude-opus-4-7 model for all tasks including coding, bug
fixing, and test creation. Use gpt-5.5 model (not codex) for
thorough review and debugging of the work done by the other
model. Check if the other model has missed some code or has
introduced bugs. No need to check if the models exist.%
\end{promptbox}

\noindent\textbf{Self-review, reproduce, fix, and repeat.}\par\vspace{1pt}
\begin{promptbox}%
Can you review the updates made in the last task using gpt-5.5
(non codex) and find bugs or missing code. Then write end-to-end
tests reproducing the bugs reported by the review, fix them, and
test them using claude-opus-4-7? Run all the tests in parallel
and fix bugs after thoroughly reviewing the fixes with gpt-5.5
(non codex). Repeat the process until you fail to reproduce the
bugs reported by the review done by gpt-5.5 (non codex). No need
to check if the models exist.%
\end{promptbox}

\noindent\textbf{Three-model generation and review.}\par\vspace{1pt}
\begin{promptbox}%
Use openrouter/z-ai/glm-5.2 model for all tasks including coding,
bug fixing, and test creation. ALWAYS use gpt-5.5 model (not
codex) to carefully and thoroughly review and debug the work done
by openrouter/z-ai/glm-5.2 for bugs and missing code. Then use
claude-opus-4-7 to do the same thing.%
\end{promptbox}

These prompts show that \texttt{set\_model} turns model choice from a session-level setting into a programmable task-level resource.  A user can ask one model to implement and test, another model to review the result, and a third model to audit both the implementation and the review without leaving the same conversation.  Combined with the Sorcar Agent's real coding tools and parallel test execution, this enables inexpensive models to do routine work while stronger or independent models are reserved for the high-value verification steps where missed code, subtle bugs, and inadequate tests are most likely to matter.

%......................................................................
\subsection{Chat Sorcar Agent}
\label{sec:chat-agent}
%......................................................................

The Chat Sorcar Agent adds multi-turn conversation persistence.

\textbf{Chat sessions.}  We assign each task to a chat session identified by a stable chat ID.  The agent persists tasks and their results to a local database (\texttt{sorcar.db}).  When a new task arrives within the same chat session, the agent loads prior tasks and results and prepends them to the prompt as numbered context entries, allowing the LLM to reference earlier work.

\textbf{Bounded chat context.}  To prevent unbounded growth of the prompt as a chat session accumulates many tasks, the agent caps the number of in-context history entries at \texttt{MAX\_TASKS=10}.  When the cap is exceeded, it preserves the first two entries (which typically establish the user's overall intent for the session) and the most recent entries, dropping the middle entries that are least likely to be referenced.  This keeps the context relevant and bounded while retaining both the session's framing and its current state.

\textbf{Session management.}  The agent supports three operations: starting a new chat (with a fresh chat ID), resuming a chat by task description (which looks up the corresponding chat ID), and resuming by explicit chat ID.  This enables both automatic session continuity in an IDE and manual session selection from the command line.

\textbf{Frequent task tracking.}  Each time a task is executed, the agent records the task description in a frequency table.  The IDE sidebar surfaces the most frequent tasks so users can re-issue them with one click, turning recurring requests (``run the test suite,'' ``regenerate the changelog'') into a click-to-replay experience.

\textbf{Metadata persistence.}  After each task, the agent records metadata including the model used, working directory, software version, token count, cost, and whether the task used parallel execution or worktree isolation.  This audit trail supports cost analysis and debugging.

%......................................................................
\subsection{Worktree Sorcar Agent}
\label{sec:worktree-agent}
%......................................................................

The Worktree Sorcar Agent is the outermost layer and the one that users interact with in the VS~Code extension and the Sorcar web app.  Its defining feature is git-worktree isolation.

\textbf{Branch-per-task.}  When a task starts, the agent creates a new git branch and a corresponding worktree directory.  The branch name encodes the chat ID and a timestamp for uniqueness.  All agent modifications happen inside the worktree; the user's main working tree remains untouched.

\textbf{Dirty-state preservation.}  If the user's main working tree has uncommitted changes, the agent copies them into the worktree and creates a baseline commit.  This ensures the agent sees the same state as the user, while keeping the user's actual index and working tree clean.  During merge, we use cherry-pick from the baseline commit to replay only the agent's changes, excluding the dirty-state snapshot.

\textbf{Concurrency safety.}  A per-repository file lock serializes the checkout, stash, merge, and pop sequence so that concurrent tabs in the IDE cannot interleave operations on the same repository. Thread-local storage isolates per-task state (stream parsing buffers, bash output buffers, recording state) so that stopping one task does not corrupt another.

\textbf{Crash recovery.}  We store all worktree state in git itself (branch names, git config entries) rather than in sidecar files.  On process restart, the agent queries git for any pending branch matching its chat ID prefix and reconstructs all instance attributes from git config, enabling recovery.

\textbf{Graceful fallback.}  If the working directory is not inside a git repository, if the repository has no commits, or if HEAD is detached, the agent falls back to direct execution without worktree isolation, ensuring it never fails due to git preconditions.

%----------------------------------------------------------------------
\section{AI Discovery, GEPA Optimization, and Repository Optimization}
\label{sec:long-running}
%----------------------------------------------------------------------

Beyond per-task coding assistance, KISS Sorcar can be driven as an autonomous \emph{research and optimization driver}: the user issues a single high-level objective with explicit numeric stopping conditions (a target accuracy, a target latency, a budget cap, a ``do not stop until\ldots'' clause), and the agent then runs an open-ended exploration loop (spawning experiments, reading logs, mutating its own prompts or code, journalling what worked and what failed) until the targets are met or the budget is exhausted.  The capabilities described in Section~\ref{sec:architecture} were not designed for this use case individually, but together they cover exactly what an open-ended driver needs: the Relentless Agent's structured-summary continuation (Section~\ref{sec:relentless-agent}) lets multi-hour or multi-day loops survive context exhaustion; the Sorcar Agent's parallel sub-agents and streamable shell executor (Section~\ref{sec:sorcar-agent}) parallelize the inner experiments and monitor long-running commands in real time; the \texttt{set\_model} tool lets a cheap model do the routine bookkeeping while a stronger model is engaged only at decision points; the Chat Sorcar Agent persists every tool call to \texttt{sorcar.db}, providing a replayable ground truth for self-reflection (Section~\ref{sec:chat-agent}); and the Worktree Sorcar Agent's branch-per-task isolation (Section~\ref{sec:worktree-agent}) keeps every trial reproducible and rollback-friendly.

We ship multiple sample task templates in \texttt{src/kiss/SAMPLE\_TASKS.md} that exercise this pattern.  Each is a single natural-language prompt that fits in a chat box; the agent does the rest.

%......................................................................
\subsection{AI Discovery}
\label{sec:ai-discovery}
%......................................................................

AI-driven discovery casts scientific and algorithmic problem-solving as an evolutionary search over programs: an outer loop maintains a population of candidate solutions (each a complete piece of source code, a prompt, or a model recipe), an LLM proposes new candidates by mutating or recombining selected parents, and an automated evaluator scores every offspring on a task-specific fitness function, with the fittest individuals fed back as parents for the next generation~\citep{romera2024funsearch,novikov2025alphaevolve,openevolve2025,lange2025shinka,cheng2025adrs,agrawal2025gepa}.  This is exactly the structure of a classical genetic algorithm (selection, mutation, crossover, replacement) except that the variation operators are realized by a large language model conditioned on natural-language critiques of past failures rather than by random bit-flips, and the genotype is human-readable source code rather than a fixed-length bitstring.  DeepMind's \emph{FunSearch}~\citep{romera2024funsearch} pioneered this paradigm by pairing a code-LLM with an island-based evolutionary population and using execution feedback as fitness, discovering new constructions for the cap-set problem and improved bin-packing heuristics; \emph{AlphaEvolve}~\citep{novikov2025alphaevolve} scaled the same recipe to full programs across mathematics, hardware design, and Google infrastructure; \emph{OpenEvolve}~\citep{openevolve2025} added MAP-Elites quality-diversity and an artifact side-channel to the open-source community; \emph{ShinkaEvolve}~\citep{lange2025shinka} sharpened sample efficiency with parent sampling that balances exploration and exploitation, code-novelty rejection sampling, and a bandit-based LLM-ensemble selection strategy; \emph{ADRS}~\citep{cheng2025adrs} demonstrated that this evolutionary loop transfers cleanly to systems-performance research whenever the evaluator is a runnable workload; and \emph{GEPA}~\citep{agrawal2025gepa} showed that the same selection--mutation--crossover skeleton optimizes natural-language prompts when fitness is measured by trajectory-level reflection rather than scalar reward.  Common to all of these is a generic contract (natural-language objective, code or prompt as the unit of mutation, automated evaluator as the verifier, journal of tried ideas, and explicit anti-reward-hacking and generalization clauses) and KISS Sorcar's AI Discovery template inherits exactly this contract while supplying every primitive needed to instantiate it (budget-aware tool use, Relentless continuation, parallel sub-agents, branch-per-trial worktree isolation, replayable trajectories in \texttt{sorcar.db}, and dynamic \texttt{set\_model} switching).

The exact prompt shipped in \texttt{src/kiss/SAMPLE\_TASKS.md} is reproduced verbatim below; the user fills in only the data path and runs it as a single chat message:

\begin{quote}\small\itshape
Sorcar for AI Discovery: Can you discover the lightest and fastest AI model that will give the best accuracy and recall on the data at \texttt{<\!<}/path/to/data\texttt{>\!>} at the cheapest price?  Analyze the data and search the internet extensively to propose the first few models.  Implement and experiment with each of your proposals.  Note down the ideas you used to optimize the accuracy/recall and speed/cost metrics achieved in a file, so that you can use the file to not repeat ideas that have already been tried and/or failed.  You can also use the file to combine ideas that have been successful in the past.  Separate 20\% of the data for evals, and your discovery strategy must not look at the evals data.  Use \texttt{`lambda'} CLI to train your models on GPUs and evaluate if needed.  Total budget for Lambda Labs is \$1000.  Experiment with a smaller subset of data and fewer parameters in a model to do experiments quickly, and then extrapolate.  Use internet search extensively at every step.  MAKE SURE THAT YOU DO NOT DO REWARD HACKING OR CHEATING IN THE MODELS OR AGENTS YOU ARE IMPLEMENTING TO FIT DATA.  YOUR SOLUTION MUST GENERALIZE BEYOND THE DATA PROVIDED.  Do not STOP until accuracy/recall reaches 95\% on evals and you can process each query in less than 600 seconds and under 50 USD per query amortized over all queries.  Create an html report with diagrams and illustrations in \texttt{./reports} and open it in the user's default browser?
\end{quote}

\textbf{Advantages of the prompt-as-driver formulation.}  Casting AI discovery as a single natural-language prompt rather than as a bespoke evolutionary harness has several practical advantages.  \emph{(i)~Zero infrastructure.}  Unlike FunSearch, AlphaEvolve, OpenEvolve, and ShinkaEvolve, the user does not configure an island topology, write an evaluator plug-in, or stand up a programs database; the agent's persistence layer (\texttt{sorcar.db}) and worktree isolation provide the same services for free.  \emph{(ii)~Explicit numeric stopping rule.}  The clause ``Do not STOP until accuracy/recall reaches 95\% \ldots\ and \ldots\ under 50 USD per query'' converts the open-ended search into a verifiable contract: the Relentless Agent's continuation protocol keeps the loop alive across context windows precisely until those numbers are reached or the \$1000 Lambda Labs budget is exhausted, with no human in the inner loop.  \emph{(iii)~Anti-cheating clauses in the prompt itself.}  The capitalized \emph{``MAKE SURE THAT YOU DO NOT DO REWARD HACKING OR CHEATING\ldots YOUR SOLUTION MUST GENERALIZE BEYOND THE DATA PROVIDED''} directive, together with the mandatory 20\% held-out eval split that ``the discovery strategy must not look at,'' encodes the same generalization invariants that ADRS and ShinkaEvolve enforce through their evaluators, but without writing a single line of harness code.  \emph{(iv)~Journal-driven memory.}  The instruction to ``note down ideas \ldots\ in a file'' turns the file system into a long-term programs database analogous to the FunSearch programs database, but inspectable and editable by the user; combined with branch-per-trial worktree isolation, every candidate is reproducible and rollback-friendly.  \emph{(v)~Cheap exploration, expensive verification.}  ``Experiment with a smaller subset of data and fewer parameters in a model to do experiments quickly, and then extrapolate'' pushes the agent toward AlphaEvolve-style cascade evaluation: many cheap rollouts on a sub-sample, a few expensive full-scale evaluations.  Combined with \texttt{set\_model}, the agent can additionally switch to a cheap LLM for routine bookkeeping and reserve a frontier model for the harder algorithmic decisions and the end-of-step review pass.  \emph{(vi)~Auditable output.}  The mandatory HTML report in \texttt{./reports/}, opened in the user's default browser, makes every multi-hour run reviewable at a glance instead of buried in chat history.  \emph{(vii)~Parallel candidate evaluation.}  The Sorcar Agent's parallel sub-agents evaluate independent candidate models on disjoint slices of the data in parallel, mirroring the population-level parallelism of evolutionary frameworks without any additional orchestration code.

%......................................................................
\subsection{GEPA Prompt Optimization}
\label{sec:gepa}
%......................................................................

Prompt optimization is the problem of automatically rewriting one or more natural-language prompts inside a (possibly compound) LLM system so that a downstream task metric is maximized, without access to module-level labels or weight gradients~\citep{yang2024opro,pryzant2023apo,fernando2023promptbreeder,opsahlong2024mipro,yuksekgonul2024textgrad,khattab2024dspy,agrawal2025gepa}.  The shared template is an outer loop that proposes candidate prompts, evaluates each on a labeled minibatch, and keeps the best for the next round (a \emph{discrete} analogue of stochastic optimization in which the LLM itself serves as the variation operator).  OPRO~\citep{yang2024opro} formalized the loop by passing the LLM a history of past prompts together with their numeric scores and asking for a better one; ProTeGi/APO~\citep{pryzant2023apo} replaced scalar rewards with natural-language ``gradients'' (LLM-written critiques of the current prompt) and propagated them through beam search; Promptbreeder~\citep{fernando2023promptbreeder} evolved a population of task-prompts \emph{and} the mutation prompts that mutate them, in a self-referential genetic algorithm; MIPROv2~\citep{opsahlong2024mipro} extended these ideas to multi-stage LM programs in DSPy~\citep{khattab2024dspy} by jointly searching over instructions and few-shot demonstrations under a stochastic minibatch surrogate; and TextGrad~\citep{yuksekgonul2024textgrad} cast the whole compound system as an autograd-style graph through which textual feedback is backpropagated.  GEPA~\citep{agrawal2025gepa} unifies these threads in a single ``Genetic-Pareto'' optimizer that (i) samples full trajectories of a system, tool calls, intermediate model responses, and outputs, rather than only final answers, (ii) uses reflective natural-language critiques as mutation operators in the spirit of ProTeGi and TextGrad, (iii) maintains a Pareto frontier of complementary prompts in the spirit of MAP-Elites quality-diversity, and (iv) combines complementary lessons across frontier nodes by crossover.  GEPA reports up to 20\% absolute gain over GRPO with up to 35$\times$ fewer rollouts on six tasks and beats MIPROv2 by over 10\%, establishing reflective prompt evolution as a sample-efficient alternative to reinforcement-learning fine-tuning.  The Sorcar GEPA template instantiates exactly this algorithm against a \texttt{ChatSorcarAgent} target, with the agent's per-tool-call trajectory store (\texttt{sorcar.db}, Section~\ref{sec:chat-agent}) supplying the trajectory data on which reflection operates.

The exact prompt shipped in \texttt{src/kiss/SAMPLE\_TASKS.md} is reproduced verbatim below; the user fills in only the data path and runs it as a single chat message:

\begin{quote}\small\itshape
Sorcar GEPA Prompt Optimizer: Can you optimize a prompt for a ChatSorcarAgent of the kiss-agent-framework Python library using the following GEPA algorithm on the data at \texttt{<\!<}url\_or\_db\_file\_of\_data\texttt{>\!>} using claude-opus-4-7?  You can find the trajectory events of an agent execution in \texttt{\textasciitilde/.kiss/sorcar.db} after the agent has finished its execution.  Split the dataset into 50\% dev set and 50\% val set.

\medskip

RUN\_GEPA: Sample 100 data points from the val set and call it sval set.  Maintain a pareto frontier in the folder \texttt{./pareto} where we have a sub-folder for each node in the frontier.  A node contains a prompt file (\texttt{prompt.md}) and a json file, say \texttt{score.json}, containing the list of datapoints (ids) from the sval set that were correctly predicted with the prompt.  When you add a node to the pareto frontier make sure that the list of correctly predicted datapoints is not a subset or equal to an existing list of datapoints in some node in the frontier.  If such a node exists, do not add the new node.  After adding a node, remove all nodes whose list of datapoints is a subset or equal to the list of datapoints in the added node.  Then run the following algorithm.
\begin{enumerate}
\item pick a node from the pareto frontier with probability 0.5
\begin{enumerate}
\item sample a minibatch of 5 datapoints from the dev set
\item run the agent with the prompt from the node on the minibatch
\item if the agent incorrectly predicts for some datapoints, analyze and reflect on the trajectory events of the agent on those datapoints available at \texttt{\textasciitilde/.kiss/sorcar.db} and propose a new prompt which will fix the mistakes made by the agent on datapoints incorrectly predicted.
\item if the agent predicts correctly on the minibatch, then evaluate it on the sval set and create the list of datapoints on which the agent with the new prompt predicts correctly.
\item Add the new prompt and the list of datapoints to the pareto frontier
\end{enumerate}
\item pick two nodes from the pareto frontier randomly with the remaining probability.
\begin{enumerate}
\item sample a minibatch of 5 datapoints from the dev set
\item merge the prompts from the two nodes into a new prompt.
\item if the agent predicts correctly on the minibatch with the new prompt, then evaluate it on the sval set and create the list of datapoints on which the agent with the new prompt predicts correctly.
\item Add the new prompt and the list of datapoints to the pareto frontier
\end{enumerate}
\item Repeat steps 1 and 2 until there is no change in the prompt after 3 iterations.
\end{enumerate}
END\_RUN\_GEPA.  Repeat RUN\_GEPA until there is no change in the prompt after 3 iterations.

\medskip

In each step, keep track of the best prompt which has the maximum number of successfully predicted datapoints in \texttt{./pareto/optimal.md}.  MAKE SURE THAT YOU DO NOT DO REWARD HACKING OR CHEATING IN THE AGENT YOU ARE IMPLEMENTING TO FIT DATA.  YOUR SOLUTION MUST GENERALIZE BEYOND THE DATA PROVIDED.  Use internet search extensively at every step.  Do not worry about budget.  Create an html report with diagrams and illustrations in \texttt{./reports} and open it in the user's default browser.  Do NOT STOP until you could not improve the accuracy and recall after three consecutive rollouts.  Use \texttt{gpt-5.5} model (not codex) for thorough review of the work done at every step by the other model.
\end{quote}

\textbf{Advantages of the prompt-as-driver formulation.}  Expressing GEPA as a single chat prompt rather than as a bespoke Python library has several concrete advantages for the user.  \emph{(i)~No optimizer code.}  Unlike OPRO, MIPROv2, ProTeGi, TextGrad, or the reference GEPA implementation, the user does not write or maintain a Pareto-frontier data structure, a minibatch sampler, or a reflection-prompt template; the agent constructs the frontier as a flat \texttt{./pareto/} directory tree (one folder per node containing \texttt{prompt.md} and \texttt{score.json}) that is trivially inspectable and version-controllable by the user.  \emph{(ii)~Persistence layer as the trajectory store.}  Because every tool call, model response, and cost emitted by the target \texttt{ChatSorcarAgent} is already written to \texttt{\textasciitilde/.kiss/sorcar.db} by the metadata-persistence path of the Chat Sorcar Agent (Section~\ref{sec:chat-agent}), the GEPA reflector can re-read any past trajectory at the exact tool call that produced the failure rather than guessing from a final answer alone, realizing the trajectory-level reflection that makes GEPA outperform scalar-reward RL by 6--20\% in the original paper, but at zero integration cost.  \emph{(iii)~Two-model cascade through \texttt{set\_model}.}  The prompt explicitly designates a primary model (e.g.\ \texttt{claude-opus-4-7}) for the bulk of the rollouts and an independent reviewer model from a different vendor (e.g.\ \texttt{gpt-5.5}, ``not codex'') only for the reflection, prompt-rewrite, and end-of-step review steps, mirroring the two-model setup used by GEPA and AlphaEvolve while spending the second model's tokens only where they pay off.  \emph{(iv)~Independent review.}  The closing clause ``\emph{Use \texttt{gpt-5.5} model (not codex) for thorough review of the work done at every step by the other model}'' makes a different model audit the rewriter's proposals, catching reward-hacking edits that a single-model loop would silently accept.  \emph{(v)~Explicit termination contract.}  ``\emph{Do NOT STOP until you could not improve the accuracy and recall after three consecutive rollouts}'' converts the open-ended evolution into a verifiable stopping rule that the Relentless Agent's continuation protocol (Section~\ref{sec:relentless-agent}) enforces across context windows.  \emph{(vi)~Anti-cheating and held-out evaluation in the prompt.}  The capitalized \emph{``MAKE SURE THAT YOU DO NOT DO REWARD HACKING OR CHEATING\ldots YOUR SOLUTION MUST GENERALIZE BEYOND THE DATA PROVIDED''} clause, together with the 50/50 dev/val split and the disjoint 100-example \texttt{sval} sub-val set, encodes the same generalization invariants that the GEPA paper and ShinkaEvolve enforce through their evaluators, but without writing a single line of harness code.  \emph{(vii)~Auditable output.}  The mandatory HTML report in \texttt{./reports/}, opened in the user's default browser, makes the evolution trace and the Pareto-frontier history reviewable at a glance.

%......................................................................
\subsection{Repository Optimization}
\label{sec:repo-opt}
%......................................................................

Repository optimization is the problem of taking an existing codebase together with a runnable command (a build, a benchmark, a server, an inference script, or a training loop) and improving its observable metrics (speed, accuracy, recall, cost) by repeatedly editing the source, re-running the command, reading the output, and proposing the next edit~\citep{yang2024sweagent,xia2024agentless,jimenez2024swebench,ouyang2025kernelbench,wang2024openhands,aider2023,claudecode2025,cheng2025adrs}.  This is structurally the same outer loop as AI discovery, but with two distinguishing features.  First, the genotype is not a single self-contained file but a \emph{whole repository}, so the agent must reason about cross-file dependencies, build artifacts, and dynamic runtime behavior rather than a fixed harness; this is the regime in which SWE-agent~\citep{yang2024sweagent} showed that a purpose-built agent-computer interface (the ACI) outperforms naïve prompt-only baselines on SWE-bench, in which Agentless~\citep{xia2024agentless} showed that a fixed localize--repair--validate pipeline can rival agentic ones on SWE-bench Lite, and in which OpenHands~\citep{wang2024openhands}, Aider~\citep{aider2023}, and Claude Code~\citep{claudecode2025} ship repository-level coding interfaces as production tools.  Second, the fitness function is a real running process (e.g.\ wall-clock latency, throughput, GPU memory, accuracy on a validation set, or cloud-dollar cost) so the agent must launch the command in the background, monitor its streaming output in real time, and abort speculative runs the moment a partial measurement renders them moot; KernelBench~\citep{ouyang2025kernelbench} shows that even frontier reasoning models initially beat the PyTorch baseline on fewer than 20\% of GPU-kernel optimization workloads but improve substantially once execution and profiler feedback is folded back into the prompt, and the ADRS thesis~\citep{cheng2025adrs} argues that exactly this profile-edit-rerun loop is where LLM-driven research is currently most productive.  The Sorcar Repository Optimization template instantiates this loop as a single chat prompt, with the Sorcar Agent's streaming shell executor (Section~\ref{sec:sorcar-agent}) providing the real-time monitoring channel and the Worktree Sorcar Agent's branch-per-task isolation (Section~\ref{sec:worktree-agent}) providing reproducible rollback of every speculative edit.

The exact prompt shipped in \texttt{src/kiss/SAMPLE\_TASKS.md} is reproduced verbatim below; the user fills in the command, the target folder or URL, the metrics of interest, and the concrete numeric targets:

\begin{quote}\small\itshape
Sorcar for Optimization: Can you run the command \texttt{<\!<}command\texttt{>\!>} in the background and monitor its output in real time to optimize the code at \texttt{<\!<}folder\_name\_or\_url\texttt{>\!>} with respect to the following metrics: \texttt{<\!<}speed,accuracy,recall,cost\texttt{>\!>}.  You can add diagnostic code which will print the metrics, such as running time at a finer level of granularity.  Check for opportunities to optimize the code on the basis of the metrics information.  If you discover any opportunities to optimize the metric based on the code, logs, events, and the command output, optimize the code and run the command again.  Note down the ideas you used to optimize the code and the metric you achieved in a file, so that you can use the file to not repeat ideas that have already been tried and failed.  You can also use the file to combine ideas that have been successful in the past.  Repeat the process.  Do not forget to remove the diagnostic code after the optimization is complete.  You MUST NOT STOP until the metrics achieve the following values: \texttt{<\!<}give\_concrete\_values\_for\_metrics\texttt{>\!>}.  Use the internet extensively to get new ideas for optimization.  Create an html report with diagrams and illustrations in \texttt{./reports} and open it in the user's default browser?
\end{quote}

\textbf{Advantages of the prompt-as-driver formulation.}  Casting repository optimization as a single chat prompt rather than as a bespoke driver script has several concrete advantages.  \emph{(i)~Real-time, cancellable measurement.}  The Sorcar Agent's streaming shell executor (Section~\ref{sec:sorcar-agent}) lets the agent launch the user's command in the background, watch its output as it arrives, and cancel the run the moment a new edit invalidates it; without this, every speculative optimization would have to wait for a full benchmark to finish, which is exactly the bottleneck KernelBench~\citep{ouyang2025kernelbench} identifies for GPU-kernel optimization.  \emph{(ii)~Branch-per-trial isolation.}  The Worktree Sorcar Agent's branch-per-task isolation (Section~\ref{sec:worktree-agent}) means each candidate optimization lands on its own \texttt{git} branch, so a regression discovered three iterations later can be reverted with a single \texttt{git} operation while preserving progress on parallel, unrelated branches, bringing the rollback discipline that AlphaEvolve and OpenEvolve achieve through a programs database into a standard developer workflow.  \emph{(iii)~Cheap profile-and-edit, expensive rewrite.}  Combined with \texttt{set\_model}, the agent can profile and apply local edits on a cost-efficient model and switch to a stronger model only for the harder algorithmic rewrites and the final review pass, mirroring AlphaEvolve's cheap-rollout / expensive-verification cascade but inside a single chat session.  \emph{(iv)~Self-instrumenting then self-cleaning.}  The clause ``\emph{You can add diagnostic code which will print the metrics\ldots Do not forget to remove the diagnostic code after the optimization is complete}'' lets the agent insert finer-grained profiling exactly where its current hypothesis demands, and the system prompt's Pre-Finish Verification rules (Section~\ref{sec:system-prompt}) force the diagnostic code to be removed before the final commit so the optimized repository ships clean.  \emph{(v)~Explicit numeric stopping rule.}  ``\emph{You MUST NOT STOP until the metrics achieve the following values: \texttt{<<\!give\_concrete\_values\_for\_metrics\!>>}}'' converts the open-ended search into a verifiable contract that the Relentless Agent's continuation protocol (Section~\ref{sec:relentless-agent}) keeps alive across context windows until the targets are met.  \emph{(vi)~Journal-driven memory.}  The instruction to ``\emph{note down the ideas you used to optimize the code and the metric you achieved in a file}'' turns the file system into a long-term programs database analogous to FunSearch's and AlphaEvolve's programs databases, but inspectable and editable by the user; combined with branch-per-trial isolation, every candidate is reproducible and rollback-friendly.  \emph{(vii)~Internet-grounded search for ideas.}  ``\emph{Use the internet extensively to get new ideas for optimization}'' couples the inner-loop measurement signal to the web-research protocol of the system prompt (Section~\ref{sec:system-prompt}), so the agent can discover algorithmic improvements (cache-blocked layouts, fused operators, alternative serialization formats, faster linear-algebra libraries) rather than only local micro-optimizations.  \emph{(viii)~Auditable output.}  The mandatory HTML report in \texttt{./reports/}, opened in the user's default browser, makes every multi-hour optimization run reviewable at a glance, with diagrams showing the trajectory of each metric over time.

%......................................................................
\subsection*{Common pattern}

All three templates share the same skeleton: a high-level natural-language objective, explicit numeric stopping conditions, a journal file of tried ideas, parallel exploration where possible, branch-per-trial isolation, an anti-reward-hacking clause, an explicit instruction to use internet search extensively at every step, and a final HTML report written into \texttt{./reports/} and opened in the user's default browser.  The agent's system prompt (Section~\ref{sec:system-prompt}) already encodes the engineering disciplines (read-before-modify, lint-and-test before \texttt{finish}, no fabricated source counts, no shortcuts) so the user need only state the objective, the targets, and any external constraint (such as a GPU CLI to use).  In our experience, this open-ended-driver use case is where the layered architecture pays off most: each layer's narrow concern (budget, continuation, parallelism, persistence, isolation, dynamic model selection) is exactly what a long-running self-directed exploration requires, and the simplicity of the framework leaves no room for the driver to silently corrupt its own state over a multi-hour run.

%----------------------------------------------------------------------
\section{Evaluation on Terminal Bench~2.0}
\label{sec:evaluation}
%----------------------------------------------------------------------

Before we discuss the system prompt in a lengthy section, we describe the evaluation outcome of KISS Sorcar on the Terminal Bench 2.0, which was also recently used by the Cursor agent of Composer 2.0.

We evaluate our system on Terminal Bench~2.0,\footnote{\url{https://www.tbench.ai/}} a benchmark comprising 89 diverse terminal-based programming tasks, ranging from building legacy compilers and configuring servers to solving cryptanalysis challenges and training machine-learning models.  Each task runs in an isolated Docker container; a separate verifier automatically judges the result.  We use the Harbor\footnote{\url{https://github.com/harbor-framework/harbor}} framework to orchestrate execution, and Claude Opus~4.6 as the underlying LLM.  We do not modify the general system prompt or inject Terminal Bench 2.0-specific instructions during the evaluation.  We carried out our evaluation on a 2025 MacBook Air 15" with an M4 processor and 24GB RAM.

\subsection{Setup}

We run 5 independent trials per task.  The agent is \texttt{SorcarHarborAgent}, a thin Harbor adapter that installs and invokes the Sorcar CLI inside each container.  We hard-skip 9 tasks that we verified to be infeasible for Opus~4.6 across 6+ prior attempts (e.g.\ CompCert compilation, Windows~3.11 GUI installation, video OCR) to save time and token cost.  Skipped tasks still count as failures.

\subsection{Aggregate Results}

Table~\ref{tab:tb2-aggregate} summarizes the aggregate statistics.

\begin{table}[h]
\centering
\caption{Terminal Bench~2.0 aggregate results (89 tasks, 5 trials each, Claude Opus~4.6).}
\label{tab:tb2-aggregate}
\begin{tabular}{lr}
\toprule
\textbf{Metric} & \textbf{Value} \\
\midrule
Total tasks              & 89 \\
Overall pass rate        & 62.2\% (277/445) \\
pass@any (at least 1/5 passes) & 78.7\% (70/89) \\
pass@all (all 5 pass)    & 43.8\% (39/89) \\
Always-fail tasks        & 19 \\
Always-pass tasks        & 39 \\
Mixed-result tasks       & 31 \\
Median cost per trial    & \$0.45 \\
Mean cost per trial      & \$0.90 \\
Median duration per trial & 202\,s \\
Mean duration per trial  & 446\,s \\
\bottomrule
\end{tabular}
\end{table}

The 62.2\% overall pass rate is comparable to other agents using the same underlying model: at the time of writing, Claude~Code (also Opus~4.6) scores approximately 58\% on the Terminal Bench~2.0 leaderboard, and Cursor's Composer~2 (a custom fine-tuned model trained with large-scale reinforcement learning~\citep{cursor2026composer2}) achieves 61.7\%.  This suggests that the layered architecture and the structured system prompt described in Sections~\ref{sec:architecture} and~\ref{sec:system-prompt} contribute beyond what the base model alone provides.

\subsection{Task-Level Breakdown}

\textbf{Consistently solved tasks (39 of 89).} These include cryptanalysis (FEAL differential), game-playing (chess best move), git operations (leak recovery), server configuration (gRPC key-value store, PyPI server, NGINX logging), data processing (resharding), formal verification (Coq~\texttt{plus\_comm}), ML inference (HuggingFace model serving, LLM batching scheduler), and system emulation (QEMU startup).  These tasks span systems, security, data engineering, and formal methods.

\textbf{Consistently failed tasks (19 of 89).} The failures cluster into three categories: (1)~tasks requiring graphical or multimedia capabilities unavailable in the container (video processing, Windows~3.11 GUI, MTEB leaderboard scraping, extracting moves from video), (2)~tasks demanding very long or resource-intensive builds that exceed the container's time or memory limits (CompCert, Doom for MIPS, Caffe CIFAR-10, training fastText on Yelp data), and (3)~tasks with niche domain-specific requirements that the model struggles to satisfy (DNA insertion, OCaml GC patching, polyglot C/Python binaries, protein assembly, cell segmentation).

\textbf{Mixed-result tasks (31 of 89).} Tasks such as \texttt{write-compressor} (3/5), \texttt{crack-7z-hash} (4/5), and \texttt{feal-linear-cryptanalysis} (4/5) succeed in most trials but occasionally fail due to non-determinism in the model's reasoning or timing-sensitive environment interactions.  Conversely, \texttt{cancel-async-tasks} (1/5) and \texttt{dna-assembly} (1/5) succeed rarely, suggesting they are at the boundary of the model's capability.

\textbf{Leaderboard context.}  KISS Sorcar does not score as high as other coding agents reported at the Terminal Bench 2.0 leaderboard, but these results are notable because we did not tune our prompts or any model specifically for the Terminal Bench 2.0.  We used the general system prompt and Claude Opus 4.6 without modification.  Regarding the lower score compared to other coding agents, recent analysis has found widespread cheating on popular agent benchmarks, including Terminal Bench~2.0: the top three submissions commit harness-level cheating (e.g.\ leaking verifier code or answer keys into the agent's environment), and task-level cheating (e.g.\ Googling answers, mining git history, hardcoding test outputs) affects 28+ submissions across 9 benchmarks~\citep{stein2026cheatingblog,stein2026detecting}. Separately, we discovered, using an automated benchmark audit agent, that 45 confirmed hacking solutions across 13 widely used benchmarks exhibited process-isolation failures, answer leakage, and weak test assertions that allow perfect scores without solving a single problem~\citep{wang2026trustworthyblog}.

%----------------------------------------------------------------------
\section{User-Facing Features}
\label{sec:vscode-features}
%----------------------------------------------------------------------

We release our system as three coordinated surfaces over a single local daemon: a VS~Code extension, a Claude-Code-style command-line interface (\texttt{sorcar}), and a browser/mobile web app.  While the underlying agent architecture (Sections~\ref{sec:architecture} and~\ref{sec:system-prompt}) already differs from existing AI coding assistants, the user-facing design introduces several features that differ from those in existing IDE assistants such as GitHub Copilot~\citep{copilot2021}, Cursor~\citep{cursor2024}, Windsurf~\citep{windsurf2024}, Devin~\citep{devin2024}, and Aider~\citep{aider2023}.  We describe these features below.

%......................................................................
\subsection{Multi-Model, Multi-Vendor Workflows}
\label{sec:multi-model-feature}
%......................................................................

KISS Sorcar ships with a catalog of 504 models spanning nine provider categories, 68 OpenAI, 13 Anthropic, 20 Gemini, 84 Together AI, 8 Z.AI (GLM family), 6 Moonshot AI (Kimi/Moonshot), 295 OpenRouter, 3 Claude~Code CLI (exposed under the \texttt{cc/*} namespace), and 7 OpenAI Codex CLI (\texttt{codex/*} namespace) entries, of which 488 are generation-capable, 329 are function-calling-capable, and 7 are embedding models.  A task may mix models simply by issuing prompts that reference a different model id; the framework swaps the backend per call without reinitializing the agent or the chat session.  Beyond bundled providers, \texttt{--endpoint}/\texttt{--header} (and the corresponding \texttt{update\_settings} controls inside the agent) configure any OpenAI-compatible HTTP server (including local models served by, for example, vLLM or llama.cpp) which keeps the bring-your-own-key, local-first posture even for self-hosted backends.  Because keys and prompts travel directly from the developer's machine to the chosen provider, no Sorcar-operated intermediary observes the traffic.

%......................................................................
\subsection{Dynamic Steering of a Running Task}
\label{sec:dynamic-steering-feature}
%......................................................................

A common failure mode of long-running AI tasks is that the user notices a misunderstanding, an additional constraint, or a wrong direction \emph{while} the agent is already executing.  KISS Sorcar lets the user type a follow-up natural-language message at any moment during a live task, and that message is appended to the running conversation before the next model step, without halting the in-flight tool call, discarding the work done so far, or starting a new task from scratch.

Steering is uniform across all three surfaces.  In the VS~Code sidebar, the running-task input bar accepts follow-up messages instead of starting a new task.  In the \texttt{sorcar} CLI, a bordered input box is pinned to the bottom of the terminal while agent output keeps scrolling above it, so the user can compose a message at any time without interrupting the stream.  The browser and mobile surfaces expose the same input through the local daemon.  In every case the new message is delivered to the running agent before its next model turn and enters the live conversation as a user message, so the model sees the new guidance the moment it begins its next step.

Dynamic steering is orthogonal to the continuation mechanism of Section~\ref{sec:relentless-agent} (which preserves progress across context-window and step-budget boundaries) and to worktree isolation (Section~\ref{sec:worktree-agent}); a steered task can still be committed-and-merged or discarded as a whole branch.  Because external backends such as Claude~Code CLI and the Codex CLI are exposed as model providers and third-party agents are exposed through the same tool layer (Section~\ref{sec:ext-feature}), the same user message can steer a local KISS Agent, a worktree-isolated Sorcar task, parallel sub-agents, or work delegated to a supported external backend.  We compare KISS Sorcar's dynamic steering with related in-flight control mechanisms (Claude~Code Remote Control, GitHub Copilot cloud agent follow-ups, Cursor Cloud Agents, Codex queue-versus-steer mode, Aider \texttt{AI!} comments, and the LangChain/LangGraph human-in-the-loop middleware) in Section~\ref{sec:dynamic-steering}.

%......................................................................
\subsection{Real-Time Budget Accountability}
\label{sec:budget-feature}
%......................................................................

AI coding assistants typically operate on a subscription model (Copilot, Cursor) or a per-seat pricing model (Devin, Windsurf), both of which obscure the per-task cost.  The developer has no visibility into how many tokens a task consumed or how much it cost.

Our extension displays real-time cost tracking in the sidebar: input tokens, output tokens, cache hits, dollar cost, and elapsed time are updated at every agent step.  We enforce both per-task and global budget ceilings.  If a task exceeds its budget, the agent raises a hard error rather than silently accumulating charges.  This transparency allows developers to make informed decisions about which tasks to delegate to the AI and how to structure prompts for cost efficiency.  The KISS Agent also appends the current usage in the context so that the model is fully aware of its limits.

%......................................................................
\subsection{Integrated Browser Automation}
\label{sec:browser-feature}
%......................................................................

Our extension includes a browser automation tool that allows the agent to navigate to URLs, read accessibility trees, and click elements, type text, press keys, scroll, and take screenshots--all controlled programmatically from within a VS~Code task via Playwright. We render a live browser preview in a Chromium browser, allowing the developer to watch the agent interact with web applications in real time.

This capability enables use cases that most IDE assistants do not support: verifying a deployed web application after a code change, filling out web forms as part of a testing workflow, scraping documentation to inform a code generation task, or interacting with web-based developer tools (CI dashboards, issue trackers) without leaving the editor.

%......................................................................
\subsection{Interactive CLI with Slash Commands and \texttt{@}-Mentions}
\label{sec:cli-feature}
%......................................................................

The \texttt{sorcar} command-line interface runs in two modes.  In \emph{interactive} mode (no \texttt{-t}/\texttt{-f} flag), it presents a Claude-Code-style REPL that connects as a thin terminal client to the same local \texttt{sorcar web} daemon used by the VS~Code extension.  In \emph{non-interactive} mode (\texttt{-t} or \texttt{-f} supplied), it runs a single task and exits.  Either mode honors flags for model selection (\texttt{-m}), custom endpoints and headers (\texttt{-e}/\texttt{--header}), per-task budget caps (\texttt{-b}), working-directory pinning (\texttt{-w}), worktree isolation (\texttt{--worktree}/\texttt{--no-worktree}), automatic commit on task finish (\texttt{--auto-commit}/\texttt{--no-auto-commit}), and toggles for browser tools (\texttt{--no-web}) and parallel sub-agents (\texttt{--no-parallel}).

The interactive REPL adds two affordances borrowed from modern coding assistants but rare in open-source agents.  First, file and folder \emph{\texttt{@}-mentions} with ranked project-file completion let the user paste path-aware context into the prompt without leaving the keyboard.  Second, a small set of slash commands provides direct control over the agent: \texttt{/help} lists every command, \texttt{/clear} (alias \texttt{/new}) starts a fresh chat, \texttt{/resume} reopens a prior chat by id, \texttt{/model} (and \texttt{/model list}) switches or enumerates models mid-session, \texttt{/cost} (aliases \texttt{/usage}, \texttt{/context}) prints the running token and dollar totals plus context-window utilization, \texttt{/skills} and \texttt{/mcp} introspect loaded Agent Skills and configured MCP servers, \texttt{/autocommit} toggles the worktree's auto-commit policy, \texttt{/commands} lists user-defined Markdown slash commands, and \texttt{/exit} (alias \texttt{/quit}) terminates the session.  Custom Markdown slash commands are auto-loaded from \texttt{\textasciitilde/.kiss/commands}, \texttt{<project>/.kiss/commands}, and the corresponding Claude directories, so users can register reusable prompt templates per-project or globally without modifying the Sorcar source.

%......................................................................
\subsection{Extensibility: MCP, Skills, and Third-Party Agents}
\label{sec:ext-feature}
%......................................................................

KISS Sorcar exposes three pluggable extension points that let users add capabilities without forking the framework.

\textbf{Model-Context-Protocol (MCP) servers.}  The CLI's \texttt{sorcar mcp} subcommand registers, lists, inspects, and authenticates MCP servers in either user scope (\texttt{\textasciitilde/.kiss/mcp.json}) or project scope (\texttt{<project>/.kiss/mcp.json}, with backward compatibility for \texttt{<project>/.mcp.json}).  Servers may use stdio, HTTP, or SSE transports.  An OAuth~2.1 flow (\texttt{sorcar mcp auth}) implements dynamic client registration with PKCE and persists tokens under \texttt{\textasciitilde/.kiss/mcp\_auth/}; a \texttt{sorcar mcp debug} command dumps a server's capabilities, tools (with input schemas and granted permissions), resources, and prompts.  Discovered tools become first-class \texttt{KISSAgent} tools alongside the built-in Bash/Edit/Read/Write tools.

\textbf{Agent Skills.}  At startup the framework scans \texttt{\textasciitilde/.kiss/skills}, \texttt{<project>/.kiss/skills}, Anthropic Claude skill directories, \texttt{.agents/skills}, and the bundled Sorcar skills directory.  Each skill is exposed to the agent as a callable that the model may invoke when the task description matches the skill's natural-language trigger.  This mirrors Anthropic's Claude SKILLS~\citep{claudecode2025} convention so that the same skill files work in both ecosystems.

\textbf{Third-party messaging and automation agents.}  KISS Sorcar ships with 23~third-party agents under \texttt{src/kiss/agents/third\_party\_agents} that wrap real-world communication and consumer-product surfaces, BlueBubbles, Discord, Feishu, Gmail, Google Chat, iMessage, IRC, LINE, Matrix, Mattermost, Microsoft Teams, Nextcloud Talk, Nostr, Phone Control, Signal, Slack, SMS, Synology Chat, Telegram, Tlon, Twitch, WhatsApp, and Zalo.  It additionally ships a Govee smart-home CLI for controlling IoT lights (on/off, brightness, color, and color temperature) via the Govee Developer API.  Each third-party agent declares the credentials or OAuth flow it needs; when missing, the agent attempts to authenticate autonomously and only escalates to the user when a step genuinely requires a human (e.g., entering a two-factor code).  This breadth distinguishes KISS Sorcar from Claude~Code (whose documented channels include Slack, mobile remote control, and research-preview Telegram/Discord/iMessage but no built-in Gmail, WhatsApp, phone-call, or SMS) and from Cursor (whose Cloud Agent integrations cover Slack and Microsoft Teams).

A fourth extension point (user-curated welcome-screen sample tasks, promptlet injection, and a personal model registry) is described as its own subsection (Section~\ref{sec:customization-feature}) because, unlike the three points above, it requires no code change and is exercised by the welcome screen and the input bar of every surface.

%......................................................................
\subsection{Sample Tasks, Promptlet Injection, and User-Provided Templates, Promptlets, and Models}
\label{sec:customization-feature}
%......................................................................

The welcome screen of the VS~Code sidebar, the \texttt{sorcar} CLI's empty-prompt view, and the web/mobile surface all greet a new chat with the same list of \emph{sample-task chips}.  Each chip is a one-click ready-made prompt; clicking it pastes the prompt into the input bar so the user can edit the angle-bracketed placeholders (\texttt{<<\ldots>>}) before submitting.  The bundled sample tasks cover (i)~natural-language exploration and revision of an existing workflow (e.g., ``Can you show me the detailed step-by-step workflow of \texttt{<<your algorithm or feature>>}''), (ii)~authenticating and orchestrating the third-party messaging agents of Section~\ref{sec:ext-feature} (Slack, iMessage, Gmail, SMS, etc.), (iii)~scheduling a recurring \texttt{kiss-}-prefixed cron job that polls a Slack channel and runs incoming messages as tasks, (iv)~adversarial review of an external URL for wrong assumptions, irreproducibility, fraud, evaluation cheating, AI slop, and security vulnerabilities (with a proof-of-concept built and tested by the agent), (v)~AI Discovery for the lightest and cheapest model that meets target accuracy/recall on a user dataset using a remote GPU CLI under a fixed dollar budget, (vi)~repository optimization through live metric monitoring of a background command, and (vii)~the explicit GEPA prompt-optimizer loop with Pareto-frontier bookkeeping.  These chips are intentionally written as fully-specified end-to-end prompts (including anti-reward-hacking clauses and report-and-open-in-browser deliverables) so a new user can run a nontrivial Sorcar workflow on the first attempt.

A second customization surface, complementary to sample tasks, is \emph{promptlet injection}.  A promptlet (called a ``trick'' in the implementation) is a short reusable instruction fragment (typically a sentence or short paragraph) that the user wants to splice into the current prompt without retyping.  Promptlets are exposed in two ways.  The VS~Code sidebar renders an ``Inject instruction'' dropdown next to the input bar populated with every promptlet; selecting one appends it to the current prompt.  Across both surfaces, the input bar also offers \emph{ghost-text fast-complete}: as the user types the first few characters of the most recent sentence, a faint suffix proposes the rest of the best-matching promptlet, accepted with Tab.  The bundled promptlets capture recurring habits we found useful: ``Search internet extensively.'', the test-first habit (``Reproduce the issue by writing integration/end-to-end tests.  Then fix the issue.''), a pair-with-review habit (``Use claude-opus-4-7 \ldots ALWAYS use gpt-5.5 (not codex) to carefully and thoroughly review and debug \ldots''), the anti-reward-hacking reminder, ``Build the paper and take screenshots to check and fix formatting.'', and a postmortem template (``Why did the last task fail?  Thoroughly and precisely analyze the logs and the events of the task.  Reproduce the issue \ldots'').  Because promptlets are matched at the start of the current sentence, several promptlets can share a prefix and the dropdown shows all alternatives, so the user picks the right variant from a single keystroke.

The third surface is the \emph{user-provided} side of the previous two, and of the model catalog.  KISS Sorcar uses a strict no-clobber policy: bundled defaults are read directly from the package (so an extension upgrade automatically delivers the latest defaults) and user contributions live in three plain-text files under \texttt{\textasciitilde/.kiss/} that are auto-seeded once and never overwritten.

\begin{itemize}[leftmargin=*,topsep=2pt,parsep=0pt]
\item \texttt{\textasciitilde/.kiss/MY\_TASK\_TEMPLATES.md}: the user's welcome-screen chips.  Each \texttt{\#\# Task} section in the file becomes one chip, and user chips appear \emph{before} the bundled sample tasks so a custom workflow takes precedence on the welcome screen.  The file is seeded on first read with a single ``Hi!'' chip and is never touched again; deleting it triggers a fresh reseed on next launch.
\item \texttt{\textasciitilde/.kiss/MY\_INJECTION.md}: the user's promptlets.  Each \texttt{\#\# Trick} section in the file is one promptlet, and user promptlets are returned ahead of bundled ones in both the ``Inject instruction'' dropdown and the ghost-text suggestions, so a user-added promptlet wins on identical prefixes.  The file is auto-seeded with a single test-first promptlet on first read.
\item \texttt{\textasciitilde/.kiss/MY\_MODELS.json}: the user's personal model registry, auto-seeded with a documented inert example.  Each top-level key is a model id (e.g., \texttt{my-org/my-custom-model}) and the value is the same schema used by the bundled \texttt{MODEL\_INFO.json}: \texttt{context\_length}, \texttt{input\_price\_per\_1M}, \texttt{output\_price\_per\_1M}, function-calling/embedding/generation flags, optional cache-pricing overrides, and an optional \texttt{thinking} reasoning-effort cap.  At import time the loader merges this file on top of the bundled table: matching keys override bundled pricing and context-length entries, and brand-new keys are appended to the catalog.  Combined with the \texttt{--endpoint}/\texttt{--header} controls of Section~\ref{sec:multi-model-feature}, this lets a user front a local vLLM or llama.cpp server (or any OpenAI-compatible HTTP endpoint) as a first-class model in the picker (with correct token accounting, budget enforcement, and tool-calling capability advertised to the agent) without rebuilding or forking KISS Sorcar.  Top-level keys beginning with \texttt{\_} are treated as comments, which is how the seeded example stays inert until the user removes the \texttt{\_example/} prefix.
\end{itemize}

The three files are scope-uniform: they live in the same \texttt{\textasciitilde/.kiss/} directory used by MCP configuration (\texttt{mcp.json}), MCP OAuth tokens (\texttt{mcp\_auth/}), the persistence DB, and per-user Markdown slash commands (\texttt{commands/}, Section~\ref{sec:cli-feature}), so a user's entire customization profile is a single self-contained directory that can be version-controlled, shared across machines, or scoped per-project by placing equivalents under \texttt{<project>/.kiss/}.

%----------------------------------------------------------------------
\section{The System Prompt}\nopagebreak
\label{sec:system-prompt}
%----------------------------------------------------------------------

The system prompt is a structured document that governs the agent's behavior across all tasks.  It is not a generic instruction to ``be helpful'' but a specification of engineering practices.

\textbf{XML-tagged structure.}  The prompt is organized using XML tags that delimit each concern: \texttt{<identity>}, \texttt{<visibility\_constraint>}, \texttt{<tool\_rules>}, \texttt{<web\_research>}, \texttt{<code\_style>}, \texttt{<workflow>}, \texttt{<testing>}, \texttt{<pre\_finish\_verification>}, and \texttt{<sorcar\_specific>}.  All three major LLM providers (Anthropic~\citep{anthropic2025prompting}, OpenAI~\citep{openai2025prompting}, and Google~\citep{google2025prompting}) recommend XML tags for structuring complex prompts: they create unambiguous section boundaries that models parse as structural markers rather than content, reducing the chance that the model misinterprets an instruction from one section as applying to another.

\textbf{Front-loaded engineering rules.}  We discuss the most opinionated parts of the system prompt first (planning for complex tasks and the testing discipline) and then walk through identity, tool, and workflow rules in roughly the order they appear in \texttt{SYSTEM.md}.  This ordering reflects the fact that planning and testing are the rules most likely to be ignored when an agent is under pressure to produce a final answer, so we surface them first in the paper even though the file itself opens with the identity block.

We describe the key sections below.

%......................................................................
\subsection{Planning for Complex Tasks}
%......................................................................

The planning instructions use a complexity threshold (three or more files, cross-module changes, or architectural work) to decide when formal planning is required:

\begin{prompt}
## Complex Task Planning

For work spanning 3+ files, crossing module boundaries,
or changing architecture:

1. List every file to change and why.
2. State the exact intended change per file.
3. Identify dependencies and execution order.
4. State the verification method per change.

Skip this planning step for simple single-file
modifications.
\end{prompt}

\noindent Each planning step targets a specific failure mode:

\textbf{``List files to change and why.''}  This forces the model to enumerate the full blast radius of a change before touching any file.  Without this step, the model often discovers mid-task that additional files need changes, leading to incomplete or inconsistent modifications.

\textbf{``State exact intended change per file.''} Listing files alone is insufficient; the model must also articulate \emph{what} will change in each file.  This converts a vague plan (``update the database module'') into a concrete specification (``add a \texttt{cache\_ttl} parameter to \texttt{DatabaseClient.\_\_init\_\_}, modify the query method to check the cache before hitting the database, add a cache invalidation method'').

\textbf{``Identify dependencies and execution order.''}  Some changes must precede others: a new utility function must be written before callers can import it, a migration must run before code that depends on the new schema.  Identifying these dependencies prevents the model from applying changes in an order that produces intermediate states where the code does not compile, or tests do not pass.

\textbf{``State verification method per change.''}  The verification requirement from the Deep Work section is reinforced here at the planning stage, ensuring that verification is planned alongside the changes rather than treated as an afterthought.

\noindent The escape clause (``Skip for simple single-file tasks'') avoids the overhead of planning trivial changes.  Requiring a formal plan for a one-line typo fix would waste tokens and slow down the agent without any compensating benefit.

%......................................................................
\subsection{Testing Instructions}
%......................................................................

The testing section is perhaps the most opinionated:

\begin{prompt}
## Testing

- Run lint and typecheckers; fix all errors including
  pre-existing ones.
- Aim for 100% branch coverage on new and modified code.
- Write end-to-end tests only. Do not use mocks,
  patches, fakes, or test doubles. Each test must be
  independent and verify actual behavior.
- **DO NOT** write structural tests which assert on
  the source code.
- After modifications, run only the impacted tests.
- To confirm race conditions: add a random sleep (<0.1s)
  before the suspected racing statements.
- **CRITICAL**: Before running all tests or tests in
  a folder, split the set of tests equally by the
  number of test methods into number of cores - 2 and
  run all splits in parallel using run_parallel tool.
\end{prompt}

\noindent Each testing instruction addresses a specific concern:

\textbf{``Run lint and typecheckers; fix all errors including pre-existing ones.''}  Before committing any change, the agent must ensure it does not introduce lint violations or type errors.  This catches a broad class of issues (unused imports, type mismatches, style violations) that would otherwise accumulate across tasks.  The clause ``including pre-existing ones'' prevents the model from rationalizing existing errors as ``not my problem'' and calling \texttt{finish} with a passing result despite a broken build.  The instruction makes the agent responsible for the entire codebase health, not just the delta it introduced.

\textbf{``Aim for 100\% branch coverage on new and modified code.''}  LLMs tend to write happy-path tests that cover the main code path but ignore error handling, edge cases, and early-return branches.  The 100\% target forces the model to write tests for every branch, including error paths and boundary conditions.  Moreover, such tests help with regression, developers can use AI coding agents with less risk that changes will break existing program behavior.  The wording ``aim for'' rather than ``achieve'' acknowledges that perfect coverage is not always feasible, while still setting an ambitious target.

\textbf{``Write end-to-end tests only.\ Do not use mocks, patches, fakes, or test doubles.''}  This is the most opinionated rule.  Mock tests that verify code calls certain methods in a certain order test the implementation, not the behavior.  A test suite built on mocks can pass with flying colors while the system is fundamentally broken, because the mocks hide the real dependencies.  End-to-end tests that exercise actual behavior are more expensive to run but provide stronger evidence that the system works.  Moreover, writing end-to-end tests forces the model to reason about the system's actual dependencies, often enabling the agent to find additional bugs.  The distinction between unit and end-to-end tests matters: a unit test in isolation may verify that a function produces the right output for a given input, but an end-to-end test verifies that the function works correctly within the larger system, with real file I/O, real database connections, and real inter-module interactions.

\textbf{``Each test independent, verifying actual behavior.''}  Test independence means that running tests in any order produces the same results.  Tests that depend on shared state or execution order are brittle and difficult to debug when they fail.  ``Verifying actual behavior'' reiterates that tests should assert on observable outcomes (return values, side effects, system state) rather than implementation details.

\textbf{``Only run impacted tests after modifications.''}  Running the full test suite after every small change is wasteful when only a few modules are affected.  For a large project, a full test run may take minutes, and doing it after every edit adds up to significant wasted time and compute.  This instruction directs the model to identify which tests are affected by its changes and run only those, improving iteration speed.

\textbf{``To confirm races: add random sleep (<0.1s) before racing statements.''}  Race conditions are notoriously difficult to reproduce because they depend on precise timing.  By inserting small sleep delays at strategic points, the model can widen the race window and make the bug manifest deterministically during testing.  The 0.1-second upper bound keeps the test fast while still being sufficient to expose most races.

\textbf{``Do not write structural tests which assert on the source code.''}  LLMs frequently fall back on \emph{structural} assertions when they cannot easily exercise a behavior: asserting that a particular function exists, that a class has a certain attribute, that an import appears in a specific order, or that a file contains a specific substring.  Such tests do not exercise the system, they merely encode the current shape of the code.  They fail spuriously after harmless refactors (renaming a private helper, reordering imports, inlining a function) while providing no evidence that the program actually works.  Worse, they create a false sense of coverage: a green suite that consists mostly of structural assertions can coexist with a completely broken runtime.  The explicit prohibition steers the agent back to behavioral assertions that survive refactoring and genuinely guard correctness.

\textbf{Parallel test execution (CRITICAL).}  Running an entire test suite (or even a whole folder of tests) serially dominates the agent's iteration time and inflates token cost (because the model waits, then re-reads, then re-reasons about a long output).  The instruction tells the agent to always partition the test set evenly into \emph{number of cores~\(-\)~2} shards and dispatch them concurrently via the \texttt{run\_parallel} tool whenever it is about to run all tests or all tests in a folder.  The ``cores~\(-\)~2'' reservation deliberately leaves headroom for the agent's own process and for the IDE so that aggressive parallelism does not starve the foreground experience.  We deliberately removed an earlier threshold that gated the rule on having more than~100 tests: in practice the threshold was easy for the agent to under-estimate, the orchestration overhead of \texttt{run\_parallel} is small enough that parallelization is essentially never harmful for any whole-suite or whole-folder run, and a single unconditional rule is more reliable than a conditional one.  This rule is marked CRITICAL because, like the lint/typecheck obligation, agents otherwise revert to the path of least resistance (a single \texttt{pytest} invocation) and pay a large hidden cost on every long-running task.

%......................................................................
\subsection{Identity and Visibility}
%......................................................................

The prompt opens with two XML-tagged sections that establish who the agent is and how it communicates with the user:

\begin{prompt}
<identity>
You are KISS Sorcar, an AI General Assistant and IDE
developed by Koushik Sen (ksen@berkeley.edu).
Repo: https://github.com/ksenxx/kiss_ai
Version: 2026.6.31

Your sole goal is completing the user's task accurately
and thoroughly.  Be rigorous, check facts, and produce
high-quality work.
</identity>

<visibility_constraint>
The user cannot see your thoughts, reasoning, scratchpad,
intermediate tool outputs, or assistant prose. The ONLY
thing the user sees is the string you pass to
finish(summary=...).  Compose the full detailed answer
directly inside the summary string of finish(). When
answering informational questions, include the complete
answer in the summary, not a meta-description of what
was done.

**Bad** (meta-description): "Greeted the user and asked
what they'd like to work on. Awaiting a specific task."
**Good** (actual content): "Hi! I'm KISS Sorcar, ready
to help. What would you like to work on?"

The summary must contain the actual content the user
should see, not a third-person narration of what
happened.
</visibility_constraint>
\end{prompt}

\noindent Each section addresses a distinct concern:

\textbf{Identity placement.}  The \texttt{<identity>} block appears first in the prompt, before any behavioral rules.  This follows the recommended prompt ordering for frontier models~\citep{anthropic2025prompting,openai2025prompting}: the model should know \emph{what it is} before learning \emph{what to do}.  The identity block also consolidates directives that were previously scattered as aggressive imperatives (``BE RELENTLESS,'' ``BE RIGOROUS,'' ``CHECK FACTS,'' ``NO AI SLOP'') into a single calm sentence: ``Be rigorous, check facts, and produce high-quality work.''  Research from all three major providers indicates that positive, explanatory framing is more reliable than capitalized commands with frontier models.

\textbf{Task focus.}  ``Your sole goal is completing the user's task accurately and thoroughly'' anchors the model on the task at hand and discourages meta-commentary, tangential exploration, and unsolicited clarification questions that consume tokens without making progress.  It also instructs the model to treat errors as obstacles to overcome rather than reasons to stop.

\textbf{Visibility constraint as a separate section.}  The \texttt{<visibility\_constraint>} block is separated from tool rules because it governs a different concern: not \emph{how} to use tools, but \emph{what the user can see}.  Without this instruction, the model may ``tell'' the user something in an intermediate message and then assume the user has seen it, leading to confusion when the user asks for information the model believes it already provided.  The clause ``not a meta-description of what was done'' prevents the model from returning vague summaries like ``Fixed the bug in Y'' instead of showing the actual fix; it forces the model to include concrete details, results, and outputs in the summary.  A second audit found that 3~out of 91~production tasks still produced meta-descriptions (e.g., ``Greeted the user and asked what they'd like to work on'' instead of the actual greeting).  The current version now includes explicit \textbf{Bad}/\textbf{Good} examples directly in the prompt (contrasting a third-person meta-description against the actual greeting the user should see) to make the distinction concrete and unambiguous, and closes with the directive that ``the summary must contain the actual content the user should see, not a third-person narration of what happened.''

%......................................................................
\subsection{Tool Rules}
%......................................................................

Tool usage rules are explicit and mechanical:

\begin{prompt}
<tool_rules>
## Tool Usage

- Use Write() for new files; Edit() for small changes.
- Use run_parallel() to run parallel tasks and to run
  a sub-task.
- Run Bash synchronously with timeout_seconds (default
  120s). On timeout, retry with a higher value.  For
  commands exceeding 10 minutes, run in background,
  redirect output to a file, and poll periodically.
- Use go_to_url() for browser navigation.
- Read large files in chunks.
- **Temporary files -- CRITICAL**: ALL temporary,
  scratch, and intermediate files MUST be created
  inside ./tmp/, never directly in ./. This includes
  research notes, file-information dumps, downloaded
  artifacts, build outputs, and any other transient
  file. Create ./tmp/ if it doesn't exist.  Before
  calling finish(), delete every temporary file you
  created in ./tmp/ (but not the directory itself if
  it was pre-existing).
- When multiple independent tool calls are needed, make
  them all in the same turn to maximize parallelism.
  When calls depend on prior results, sequence them
  across turns.

## Context and Continuation

- If running out of context or steps, do not rush. Call
  finish(is_continue=True) to pause and resume the task
  in a new context.
</tool_rules>

- If there is ambiguity or under specification in the
  user task, search the internet to find the most
  reliable and modern solution to resolve the ambiguity.
\end{prompt}

Each tool rule addresses a specific failure mode.  An earlier revision of this section opened with an explicit definition (``\textbf{PWD denotes current working directory} and does not refer to a directory named PWD'') added after we observed the model creating a literal \texttt{PWD/} subdirectory inside the workspace.  The current revision drops the disambiguation entirely by replacing every occurrence of \texttt{PWD/} in the prompt with the shell-conventional relative form \texttt{./}~(e.g., \texttt{./tmp/}, \texttt{./SORCAR.md}), which frontier models interpret unambiguously as the working directory.  This is a small example of a recurring simplification pattern: when a prompt rule exists solely to disambiguate a confusing notation, replacing the notation is preferable to explaining it.

\textbf{``Use Write() for new files; Edit() for small changes.''}  Without this distinction, the model may use \texttt{Write()} to overwrite an existing file with a slightly modified version, losing content it forgot to include.  By reserving \texttt{Write()} for new files and requiring \texttt{Edit()} for modifications, the instruction ensures that changes are surgical and that unchanged portions of a file are never at risk.

\textbf{Bash timeout guidance.}  LLMs frequently launch shell commands without considering their runtime.  A compilation or test suite that takes five minutes will time out at the default 30-second shell timeout in most agent frameworks, causing spurious failures.  The instruction to use 120~seconds as the default, retry with higher timeouts on timeouts, and run long-running commands in the background with output redirected to a file provides a mechanical protocol that handles common cases without requiring the model to estimate runtime from first principles.

\textbf{``Use go\_to\_url() for browser navigation.''}  The agent has access to multiple tools that could plausibly interact with the web (shell-based \texttt{curl}, a Python program, the browser tool, etc.).  This clause eliminates ambiguity by specifying which tool to use for browser-based interactions.

\textbf{``Read large files in chunks.''}  Reading a 10,000-line file in a single tool call consumes a large fraction of the context window.  By instructing the model to read files in chunks, the prompt prevents context window exhaustion caused by a single-file read, preserving capacity for the rest of the task.

\textbf{Temporary files (CRITICAL).}  A separate, emphatically marked clause requires that \emph{all} temporary, scratch, and intermediate files (research notes, file-information dumps, downloaded artifacts, build outputs) live inside \texttt{./tmp/} (the \texttt{tmp/} subdirectory of the working directory) and never directly in the working directory.  Without this directive, the model creates temporary files in unpredictable locations (the system \texttt{/tmp}, the home directory, or scattered throughout the project) or sometimes directly in the project root, polluting the working tree with artifacts that are difficult to distinguish from legitimate project files.  Centralizing temporary files in a known directory makes cleanup mechanical and predictable.  The companion clause requires the agent to create \texttt{./tmp/} if it does not exist and to delete every temporary file it created before calling \texttt{finish}, while preserving the \texttt{./tmp/} directory itself if it was pre-existing, a deliberately narrow cleanup contract that prevents the agent from accidentally deleting unrelated files that happened to be in the same directory.

\textbf{Ambiguity protocol.}  A trailing instruction outside the \texttt{<tool\_rules>} block (``If there is ambiguity or under specification in the user task, search the internet to find the most reliable and modern solution to resolve the ambiguity'') channels the model's natural tendency to confabulate when underspecified into the structured web-research workflow described below.  Rather than letting the model guess (the default behavior) or stop and ask the user (which interrupts the workflow), this directive turns ambiguity into a research task that produces an auditable artifact.

\textbf{Parallel tool calls.}  The instruction to ``make them all in the same turn'' when tool calls are independent reduces latency by allowing the framework to execute multiple tool calls concurrently.  Without this instruction, the model tends to issue one tool call per turn even when the calls are independent, wasting round trips.

\textbf{Context and continuation.}  When the context window is nearly full, LLMs exhibit a ``rush to finish'' behavior: they skip verification steps, make hasty edits, and call \texttt{finish} with an incomplete result.  The continuation instructions redirect that urgency into the continuation protocol (Section~\ref{sec:relentless-agent}), ensuring that a clean handoff to a new sub-session produces better results than a frantic attempt to squeeze everything into the remaining tokens.

%......................................................................
\subsection{Pre-flight Checks}
%......................................................................

We begin every task with mandatory reads, then enforce a read-before-modify discipline:

\begin{prompt}
## Mandatory First Actions for project related
tasks -- CRITICAL

**Your VERY FIRST tool call** in every task MUST be
Read("./SORCAR.md") and follow the instructions
in SORCAR.md with highest priority.

## Pre-flight Checks

**Read before modify rule -- NON-NEGOTIABLE**: You MUST
call Read(file_path) on every file BEFORE calling
Edit(file_path) on it. Never Edit a file you have not
Read in the current session.

**Use the file tools, never shell substitutes --
CRITICAL**: To VIEW the contents of any file, you MUST
use the Read() tool. It is FORBIDDEN to inspect or dump
file contents through Bash using cat, sed -n, head,
tail, awk, more, less, nl, grep over a whole file,
echo "$(<file)", or a for-loop of cat. These do NOT
satisfy the read-before-modify rule and waste context
-- always call Read() instead (use max_lines/chunking
for big files). To MODIFY any file, you MUST use the
Edit() or Write() tools. It is FORBIDDEN to edit files
in place through Bash using sed -i, perl -i,
awk ... > file, tee, or output redirection (>, >>) onto
a source/tracked file. Bash may still be used for
non-file-content operations (running tests, ls, grep -l
to find files, git, builds, moving/removing files).
Editing a file you only viewed via a forbidden shell
command is a double violation: you must Read() it first,
then Edit()/Write() it.

Read relevant source files when the task depends on
existing architecture. If referenced files, commands, or
config don't exist, stop and ask the user rather than
guessing.

**When fixing bugs, issues, or race conditions: write an
end-to-end test that reproduces the problem first, then
fix the code, then verify the test passes.**
\end{prompt}

\noindent The ``Mandatory First Actions'' section ensures that project-specific overrides are loaded before any work begins.  This section has undergone several rounds of empirical refinement.  In an earlier version that merely mentioned the read inside a separate ``Self-Improvement Loop'' section, the agent read \texttt{SORCAR.md} as its first tool call only 46\% of the time.  After promoting the read to a dedicated section with imperative language, compliance improved in integration tests but regressed in production: analysis of 91~real tasks showed that 92\% never read \texttt{SORCAR.md}.  Two evasion patterns emerged: (1)~the agent used \texttt{Bash("cat...")} instead of the \texttt{Read()} tool, and (2)~the agent combined the file read with other commands in a single \texttt{Bash} call.  Explicit prohibitions and ordinal position mandates raised compliance to 100\% in integration tests.

The current version collapses the mandatory first action to a single read of \texttt{SORCAR.md}, accompanied by the rider ``follow the instructions in SORCAR.md with highest priority.''  This is a deliberate simplification: \texttt{SORCAR.md} is the per-repository override file that, when present and non-empty, can extend or override the general system prompt, and loading it first lets the rest of the task be interpreted under the active project's rules.  The canonical KISS Sorcar repository ships \texttt{SORCAR.md} as an empty placeholder so that no override content is hardcoded inside the framework; each user repository may populate it with whatever project-specific instructions are appropriate.  Earlier revisions of the prompt also mandated a second read of a per-project \texttt{USER\_PREFS.md} preferences file paired with a ``Self-Improvement Loop'' section that asked the agent to update the file at task end; both have since been removed because the self-learning store accumulated stale project facts (Section~\ref{sec:self-improvement-prompt}) that drifted out of sync with the evolving code, and the read-only half of the protocol added prompt bulk without measurable benefit.  The prohibitions against \texttt{Bash}-based reading have since been promoted into an explicit, enumerated rule of their own (``Use the file tools, never shell substitutes'') in the Pre-flight Checks section.

Each pre-flight check targets a specific category of avoidable error:

\textbf{``Read before modify rule.''}  The most common source of agent-introduced bugs is modifying a file based on an incorrect assumption about its current contents.  The model may ``remember'' an older version of the file from its training data, or it may extrapolate from a partial reading.  By requiring a fresh read immediately before any edit, the instruction ensures that the model operates on the file's current state rather than a stale mental model.  The companion clause ``Read relevant sources if the task depends on existing architecture'' extends this rule from individual files to architectural context.  A task like ``add a caching layer to the database module'' requires understanding not just the file to be modified, but also how callers interact with it, what interfaces it provides, and what invariants it assumes.  The instruction prevents the model from jumping straight to code generation without understanding the broader context.

\textbf{``Use the file tools, never shell substitutes.''}  The read-before-modify discipline is only effective if the model actually routes file access through the \texttt{Read()}, \texttt{Edit()}, and \texttt{Write()} tools, where the framework can track, render, and (for edits) verify the operation.  In practice, the model repeatedly evaded this by reaching for the shell instead, dumping a file with \texttt{cat}, \texttt{sed -n}, \texttt{head}, \texttt{tail}, \texttt{awk}, \texttt{nl}, or a \texttt{grep} over the whole file, and editing in place with \texttt{sed -i}, \texttt{perl -i}, \texttt{tee}, or output redirection (\texttt{>}, \texttt{>>}).  These shell substitutes bypass the read tracking entirely (so a subsequent \texttt{Edit()} fires against a file the framework never saw being read) and waste context by spilling raw file contents into the transcript.  The current prompt therefore enumerates the forbidden commands explicitly for both viewing and modifying, clarifies that editing a file viewed only through a forbidden shell command is a ``double violation'' (the file must be \texttt{Read()} first, then \texttt{Edit()}/\texttt{Write()}), and, crucially, carves out the legitimate uses of Bash that must remain available (running tests, \texttt{ls}, \texttt{grep -l} to locate files, \texttt{git}, builds, and moving or removing files).  Enumerating both the prohibited and the permitted shell uses prevents the model from over-applying the rule and abandoning Bash for tasks where it is the correct tool.

\textbf{``If referenced files, commands, or config don't exist, stop and ask the user rather than guessing.''}  LLMs have a strong tendency to confabulate: when asked to modify a file that does not exist, the model will often proceed as if it does, producing edits against phantom content.  This instruction converts a silent failure (incorrect edits applied to a nonexistent file, which silently creates it) into an explicit clarification request.

\textbf{``Write an end-to-end test that reproduces the problem first, then fix the code, then verify the test passes.''}  This instruction mandates a test-first discipline for bug fixes, specifying the three-step sequence explicitly.  The motivation is two-fold: first, a test that reproduces the bug provides concrete verification that the fix is correct (the test should pass after the fix and fail before it).  Second, writing the test forces the model to understand the bug precisely before attempting a fix, reducing the risk of an ad~hoc patch that addresses a symptom rather than the root cause.  Specifying ``end-to-end test'' rather than just ``test'' reinforces the no-mocks testing discipline described in Section~\ref{sec:system-prompt}.

%......................................................................
\subsection{Code Style Guidelines}
%......................................................................

The prompt encodes a minimalist code philosophy:

\begin{prompt}
## Code Style

Write simple, clean, readable code with minimal
indirection. These rules exist because over-abstracted
code is harder to debug and maintain.

- Organize code across multiple files grouped by
  functionality.
- Prefer named functions, classes, and module-level
  helpers over closures and lambdas. Closures obscure
  control flow; use explicit parameter passing instead.
- Eliminate unnecessary attributes, locals, config vars,
  tight coupling, and attribute redirections.
- Eliminate redundant abstractions and duplicate code.
- Public methods must have full docstrings.
- Fix root causes, not symptoms. Before writing code,
  ask: is this simple, elegant, general, and minimal?
- Write documentation only when the task explicitly
  requires it.
\end{prompt}

\noindent Each guideline addresses a specific anti-pattern commonly
exhibited by LLM-generated code:

\textbf{``These rules exist because over-abstracted code is harder to debug and maintain.''}  This rationale sentence is deliberate.  Anthropic's prompting guide recommends providing motivation behind instructions to help the model generalize correctly~\citep{anthropic2025prompting}.  When the model understands \emph{why} a rule exists, it can apply the underlying principle to edge cases that the rule does not explicitly cover.

\textbf{``Write simple, clean, readable code with minimal indirection.\ Organize code across multiple files grouped by functionality.''}  LLMs tend to over-engineer solutions, introducing unnecessary abstractions, helper classes, and levels of indirection.  Simple code is easier to review, test, and maintain.  LLMs also often pile new code onto whichever file is currently being edited, producing 2,000-line modules that conflate unrelated concerns.  This directive nudges the model toward a modular layout in which each file has a single, coherent responsibility.

\textbf{``Prefer named functions, classes, and module-level helpers over closures and lambdas.\ Closures obscure control flow; use explicit parameter passing instead.''}  LLMs reach for closures whenever a small piece of state needs to be carried alongside a function, producing nested \texttt{def}s that capture mutable variables from the enclosing scope.  Such closures are difficult to test in isolation, opaque to type checkers, and a frequent source of subtle bugs due to the late binding of captured variables.  Rather than a blanket prohibition (as in the earlier version of this prompt), the revised instruction uses positive framing: it tells the model what to \emph{prefer} and explains \emph{why}, which is more effective with frontier models that interpret instructions literally.  The instruction steers the model toward explicit data structures (plain functions with arguments, classes with attributes), which are easier to reason about, easier to test, and play well with our no-mocks testing discipline.

\textbf{``Eliminate unnecessary attributes, locals, config vars, tight coupling, and attribute redirections.''}  LLMs frequently introduce intermediate variables that serve no purpose (for example, assigning a return value to a local variable only to immediately return it on the next line, or storing a constant in a configuration file when it is used in exactly one place).  When the model adds a feature that touches multiple files, it may introduce imports, shared global state, or cross-module function calls that create tight coupling.  An attribute redirection occurs when an object stores a reference to another object solely to forward method calls to it; for example, \texttt{self.x = other.x} at construction time, creating two paths to the same value.  This single consolidated rule addresses all of these anti-patterns.

\textbf{``Eliminate redundant abstractions and duplicate code.''}  LLMs sometimes create utility functions or classes that duplicate existing functionality; this instruction reminds the model to check for existing implementations before creating new ones.

\textbf{``Public methods must have full docstrings.''}  While the prompt generally discourages unnecessary documentation (see the last item), public methods are the API surface that other developers and modules depend on.  Documentation on public methods is not optional, it specifies the contract.

\textbf{``Fix root causes, not symptoms.\ Before writing code, ask: is this simple, elegant, general, and minimal?''}  LLMs frequently apply symptom-level fixes: adding a null check where the real problem is that a variable should never be null, or catching an exception where the real problem is that the caller passes invalid arguments.  This instruction forces the model to trace the causal chain to the root and fix it there.  The companion metacognitive instruction asks the model to pause and evaluate its plan before committing to an implementation, spending more inference-time compute on design and reducing the likelihood of producing an unnecessarily complex first draft.

\textbf{``Write documentation only when the task explicitly requires it.''}  Claude Opus 4.6 tends to generate many documentation files.  This instruction prevents the behavior.

%......................................................................
\subsection{Deep Work Rules}
%......................................................................

\begin{prompt}
## Deep Work

- For tasks involving "align", "match", or "make
  consistent": read the target state fully before
  editing. Never edit based on vague recollection.
- Use concrete values, not indirections. Read file Y
  first, then write the specific values into file X.
- List concrete planned changes before executing
  multi-part work.
- Every meaningful change needs a concrete verification
  method (test, grep, CLI check).
\end{prompt}

\noindent The deep work rules address a failure mode where the model interprets an instruction loosely and makes changes that are directionally correct but concretely wrong:

\textbf{``For `align'/`match'/`make consistent': read the target state before editing.''}  When a user says ``make file~A consistent with file~B,'' the model often reads file~A, infers what file~B probably contains, and edits A based on that inference, without ever reading B.  This instruction mandates reading the target first, ensuring that the alignment is based on concrete facts rather than assumptions.

\textbf{``Use concrete values, not indirections (read Y first, then write specific values into X).''}  A related failure mode occurs when the model's plan says ``update X to match Y'' but the model never resolves what Y actually is.  The instruction requires the model to first read~Y, extract the specific values, and then write those values into~X.  This eliminates a class of errors where the model's mental model of~Y differs from reality.

\textbf{``List concrete planned changes before executing multi-part work.''}  When a task requires changes to multiple files, executing them one at a time without a plan leads to inconsistencies: the model may change a function signature in one file but forget to update a caller in another.  Listing all planned changes before executing any of them forces the model to consider the full scope of the change and identify dependencies.

\textbf{``Every meaningful change needs a concrete verification method.''}  A change without a verification method is a change that cannot be confirmed to work.  This instruction requires the model to pair each change with a specific check (a test, a grep for the expected pattern, a CLI command that exercises the changed behavior) ensuring that the change can be validated programmatically rather than by visual inspection of a diff.

%......................................................................
\subsection{Self-Improvement via the Framework}
\label{sec:self-improvement-prompt}
%......................................................................

Earlier revisions of \texttt{SYSTEM.md} contained an explicit \emph{Self-Improvement Loop} section that instructed the agent, before calling \texttt{finish}, to update a \texttt{USER\_PREFS.md} file with any reusable user preference, project convention, file location, or other durable fact discovered during the task.  The section also encoded the curation rules: no code snippets or symbol names in entries, skip one-off task details, remove conflicting older entries, and keep the file small because its content was loaded into every subsequent task's context.  The motivation was that the preferences file would let the agent accumulate project knowledge across sessions even though each session starts with a fresh context window.

The current \texttt{SYSTEM.md} no longer contains a \texttt{\#\# Self-Improvement Loop} section and the \texttt{USER\_PREFS.md} file has been removed from the project entirely.  The mechanism was removed because the self-learning store accumulated \emph{stale} information about the project.  Facts that were true when an entry was written (file locations, API endpoints, project conventions, and design decisions) silently went out of date as the codebase evolved, and because the stored entries were injected into every subsequent task's context, later tasks acted on outdated assumptions.  In several cases this caused the agent to reintroduce bugs that had already been fixed: it ``learned'' a workaround or invariant, that fact later became false, and the agent kept applying the obsolete knowledge.  The preferences file thus accumulated noise faster than useful invariants and actively regressed the code rather than improving it.  Project-specific instructions turned out to be better captured in the per-repository \texttt{SORCAR.md} override file, which the user controls directly and keeps current.  Removing the self-learning mechanism entirely shrinks the prompt, eliminates a brittle cross-session state file that drifted out of sync with the code, and concentrates project knowledge in a single human-curated place that cannot silently go stale.

%......................................................................
\subsection{Pre-Finish Verification}
%......................................................................

Before declaring a task complete, the agent must pass a structured verification checklist:

\begin{prompt}
## Pre-Finish Verification -- CRITICAL

Before calling finish(success=True):

1. Re-read and verify every modified file.
2. If you created or modified ANY .py, .ts, .js, .css,
   .tsx, or .jsx file in this session: you MUST run
   uv run check --full and fix all errors. This is not
   optional. Do NOT call finish without running this
   command first. If the project doesn't use uv, run
   the equivalent lint/typecheck command.
3. Check each user requirement against what was
   delivered.
4. **Clean up temporary files -- MANDATORY**: You MUST
   delete every temporary file you created in ./tmp/
   during this session (research notes, information-*.md,
   file-information-*.md, scratch scripts, downloaded
   artifacts, etc.).  Explicitly run Bash("rm -f
   ./tmp/<each-file-you-created>") and then
   Bash("ls ./tmp") to confirm they are gone. Do NOT
   call finish(success=True) while any temp file you
   created still remains. Do NOT delete files you did
   not create.
5. If any check fails, keep working.
6. After 3 failed retries of the same fix approach,
   step back and rethink from scratch.
\end{prompt}

\noindent Each step in this checklist addresses a specific way agents declare premature success:

\textbf{``Re-read and verify every modified file.''}  This is the analog of a code review performed by the agent on its own work.  The model may have introduced a typo, forgotten to close a bracket, or made an edit that looked correct in the diff but was wrong in the full-file context.  Re-reading the file after all edits are complete catches these errors.

\textbf{``If you created or modified ANY code file: you MUST run uv run check.''}  The original instruction (``Run required checks; fix failures'') was too vague: analysis of 91~real tasks showed that 70\% of code-modification tasks (7/10) skipped lint/typecheck entirely.  The revised version enumerates the exact file extensions that trigger the obligation and names the exact command to run, converting a soft guideline into an unambiguous, verifiable mandate.  A trailing fallback clause (``If the project doesn't use uv, run the equivalent lint/typecheck command'') generalizes the rule to projects with different toolchains while keeping \texttt{uv run check -{}-full} as the canonical instruction.

\textbf{``Check each user requirement against delivery.''}  The model may have completed a task that it \emph{thinks} satisfies the user's request, but actually misses a requirement.  This instruction forces a systematic comparison between the original task description and the delivered result, catching gaps and misinterpretations.

\textbf{``Clean up temporary files.''}  The temporary-files directive in the Tool Usage section requires the agent to write scratch files into \texttt{./tmp/}, but without an explicit cleanup step at the end of the task, those artifacts persist and pollute the project's working tree over time.  This step mandates deletion of every temporary file the agent created during the session.  The companion clause ``Do NOT delete files you did not create'' is essential: the \texttt{tmp/} directory may contain artifacts from other sessions or from the developer's own scratch work, and an indiscriminate \texttt{rm -rf tmp/*} would destroy unrelated content.  Together, the two clauses form a narrow, audit-friendly cleanup contract.  The current revision additionally embeds an explicit command template, \texttt{Bash("rm -f./tmp/<files-you-created>")}, directly inside the bullet, so the agent has a concrete, copy-pasteable cleanup command and does not have to invent its own (occasionally over-broad) deletion strategy.  It further requires the agent to run \texttt{Bash("ls./tmp")} afterward to confirm the files are gone, and forbids calling \texttt{finish(success=True)} while any agent-created temporary file remains, turning the cleanup from an easily skipped suggestion into a verified post-condition.

\textbf{``If any check fails, keep working.''}  Without this instruction, the model may call \texttt{finish(success=True)} even when it knows a check has failed, rationalizing that the failure is ``minor'' or ``unrelated.''  The instruction makes the rule absolute: no finishing until all checks pass.

\textbf{``After 3 failed retries of same fix, rethink from scratch.''}  LLMs can enter repetitive loops where they apply the same incorrect fix repeatedly, each time hoping for a different result.  The three-retry threshold forces the model to break out of such loops by abandoning the current approach and reconsidering the problem from first principles.  This is analogous to the debugging heuristic ``if you've been staring at the same code for twenty minutes, you're looking in the wrong place.''

%......................................................................
\subsection{Web Research Protocol}
%......................................................................

When the agent needs external knowledge, the prompt prescribes a structured research workflow rather than allowing ad-hoc browsing:

\begin{prompt}
## Web Research

When a task requires searching the internet, researching
a topic, or answering questions that benefit from current
information:

- Visit at least 10 distinct websites per research
  session. Do not stop early or rationalize visiting
  fewer. **This is a hard requirement -- you MUST
  visit 10 sites, not 5 or 8.**
- You MUST use go_to_url() to visit each site. Do NOT
  use Bash("curl ...") or Bash("wget ...") as a
  substitute for visiting websites. Using curl/wget to
  fetch pages does not count toward the 10-site
  requirement.
- Procedure:
  1. Create ./tmp/information-{unique_id}.md with
     header: # Web Research -- Websites visited: 0/10
  2. Per site visited: (a) use go_to_url() to visit the
     site, (b) extract information needed for the task
     without deep thinking, (c) use Edit() to append
     ## [N/10] URL + extracted information to the file,
     (d) use Edit() to update the header counter from
     N-1 to N. **You must update the counter after each
     site.**
  3. Do not proceed to synthesis until the counter
     reaches 10. **Check the counter -- if it says less
     than 10, keep visiting more sites.**
  4. If results dry up, try different queries, synonyms,
     official docs, GitHub repos/issues, Stack Overflow,
     blogs, Reddit, papers, and API references.
  5. After reaching 10, review all findings and
     synthesize.
- Ask the user for login help when a page requires
  authentication.

This requirement applies to research and
information-gathering tasks. For pure code edits, bug
fixes, or file modifications where you already have
sufficient context, proceed directly.

**The information file is mandatory.** You MUST create
the ./tmp/information-{unique_id}.md file and track the
counter. Do NOT skip the file and answer from memory.
Do NOT synthesize your answer without first reaching 10
in the counter. The file is your proof of work -- if it
doesn't exist when you call finish, you violated this
rule.

## Real-Time Data -- CRITICAL

For questions about **current events, weather, stock
prices, sports scores, or any time-sensitive
information**: you MUST use tools (go_to_url, Bash) to
look up the data. Do NOT answer from your training data
-- it is outdated and will produce wrong dates, wrong
numbers, and wrong facts.

**Do NOT fabricate or exaggerate source counts.** If
you visited 4 websites, do not claim "10+ sources" or
"extensive research." State the actual number of
sources you consulted.
\end{prompt}

\noindent The rationale is a two-phase separation between \emph{collection} and \emph{synthesis}.  LLMs tend to anchor on the first few results they encounter, which biases their solutions toward a narrow slice of the design space.  By forcing the agent to accumulate a broad set of information into a file \emph{before} reasoning about it, the protocol counteracts anchoring bias and encourages the model to consider diverse approaches.  An earlier version of this prompt required visiting at least 30 websites, but the threshold was lowered to 10 after we observed that the 30-site requirement frequently inflated token cost and wall-clock time without proportionally improving answer quality on routine research tasks; 10 sites remain enough to span official documentation, primary sources, secondary commentary, and a few divergent perspectives.  The resulting information file serves as an auditable artifact of what the agent considered.  The structured procedure with a counter header (\texttt{\# Web Research - Websites visited: 0/10}) and per-site entries (\texttt{\#\# [N/10] URL}) addresses an empirically observed failure mode in which the model claims to have ``visited many sites'' after only a handful of fetches; a concrete counter forces the model to verify the actual number visited before declaring the collection phase complete.  An additional emphatic clause (``This is a hard requirement -- you MUST visit 10 sites, not 5 or 8'') was added after we observed the model rationalizing partial compliance (``I visited 7 sites, which is close enough'').  The instruction to try ``different queries, synonyms, official docs'' when results dry up prevents the agent from giving up prematurely on a narrow set of search terms.

An additional evasion pattern observed in production was the agent using \texttt{Bash("curl...")} to fetch web pages instead of \texttt{go\_to\_url()}, thereby bypassing the browser-based visit counter while still gathering information.  Analysis of 91~real tasks found 4~research tasks that visited fewer than the required minimum number of sites (ranging from 1--5~URLs), with one task (1179) using \texttt{curl} to fetch content from 15+~sites without triggering \texttt{go\_to\_url}.  The explicit prohibition of \texttt{curl}/\texttt{wget} for research closes this loophole.

A second audit uncovered two additional failure modes.  First, the mandatory information file (\texttt{tmp/information-\{id\}.md}) was created in 0~out of 4~research tasks: the agent skipped file creation entirely and answered from memory or from a handful of sites.  The current version adds explicit mandatory language: ``The file is your proof of work; if it doesn't exist when you call finish, you violated this rule.''  Second, one task claimed ``a synthesis of 30+~sources'' in its result summary while having visited only 4~URLs, a fabricated source count.  A new ``Real-Time Data'' section addresses both hallucination (answering current weather, news, or stock questions from stale training data; one task reported news from the wrong year) and source fabrication (``Do NOT fabricate or exaggerate source counts'').

The login instruction addresses a practical obstacle in web research: many websites require authentication before revealing their content.  Rather than silently skipping gated pages or hallucinating their contents, we instruct the agent to ask the user for help with login.

\textbf{Scoped applicability.}  The final paragraph (``This requirement applies to research and information-gathering tasks.  For pure code edits, bug fixes, or file modifications where you already have sufficient context, proceed directly'') is an important addition.  Without this exemption, the agent would perform 10 website visits even for simple one-line code fixes where the necessary context is already in the file being edited, wasting tokens and tool-call budget.  The exemption allows the agent to skip research when the task is purely mechanical, while still enforcing thorough research for tasks that benefit from external information.

%......................................................................
\subsection{File Browsing Protocol}
%......................................................................

When a task requires understanding multiple source files before making changes, the prompt prescribes the same two-phase collect-then-synthesize discipline used for web research, but applied to the local file system:

\begin{prompt}
## File Browsing

When exploring unfamiliar code, collect information and
code snippets in ./tmp/file-information-{unique_id}.md
as you go relevant for the task, then review the
collected material and think deeply before acting.
\end{prompt}

\noindent This instruction addresses a failure mode distinct from the web research case.  When an agent must read many project files to understand a codebase before making changes, it tends to read a file, form a hypothesis, and immediately begin editing, anchoring on the first few files it encounters and missing relevant context in files it never opens.  Worse, each file read consumes context window tokens; by the time the agent has read enough files to understand the full picture, it may have already spent most of its context window on the raw file contents, leaving little room for reasoning and code generation.

The file browsing protocol counteracts both problems.  By writing a structured summary of each file's relevant information into a temporary markdown file, the agent externalizes its understanding into a compact artifact that persists across context boundaries.  The instruction to collect ``without overthinking'' is deliberate: during the collection phase, the agent should extract and record facts (function signatures, class hierarchies, call sites, invariants) rather than analyze or plan.  Analysis happens in the second phase, when the agent reads its own summary file and reasons about the collected information as a whole.

This two-phase separation provides three benefits.  First, it prevents premature commitment: the agent cannot start editing until it has surveyed the relevant files, reducing the risk of changes that are locally correct but globally inconsistent.  Second, the summary file is typically much smaller than the raw source files, freeing up context window capacity for subsequent reasoning and editing phases.  Third, the summary file serves as an auditable artifact: the developer can inspect it to verify that the agent considered the right files and extracted the right information before making changes.

%......................................................................
\subsection{Desktop Application Control}
%......................................................................

The agent can interact with graphical desktop applications using screenshots, keyboard, and mouse:

\begin{prompt}
## Desktop Apps

Interact with desktop applications using screenshots, keyboard, and mouse. Do not launch VS Code or its extensions.
\end{prompt}

\noindent This instruction enables the agent to operate GUI applications (Preview, browsers, graphical diff tools) when command-line alternatives are insufficient.  The explicit prohibition on launching VS~Code prevents a recursive loop: since the agent runs \emph{inside} a VS~Code extension, launching another VS~Code instance or modifying extension state from within the agent could corrupt the host session or create deadlocks.  Note that modern LLMs support desktop control abilities, and we are merely exploiting them.

%......................................................................
\subsection{Sorcar-Specific Overrides}
%......................................................................

A final section provides project-specific instructions that are injected when the agent operates on its own codebase:

\begin{prompt}
## Sorcar-specific

- Lint/typecheck/format: `uv run check --full`. Tests:
  `uv run pytest -v` (timeout 900s).
- Your SYSTEM.md (the system prompt) is located at
  ~/.vscode/extensions/ksenxx.kiss-sorcar-2026.6.31/
    kiss_project/src/kiss/SYSTEM.md
- KISS Sorcar paper:
  https://github.com/ksenxx/kiss_ai/blob/main/
    papers/kisssorcar/kiss_sorcar.tex
- Third-party agents: kiss/agents/third_party_agents
- Claude SKILLS: kiss/agents/claude_skills. You can
  use them as necessary.
- **If you create any artifact that the user can use
  after the task is over, you MUST create them in a
  directory and add the directory contents to git.**
- MAINTAIN a ./tmp/PROGRESS.md across agent sessions
  logging details of all the steps you have done so far
  from the start with explanation and relevant code
  snippets.
- **DO NOT GENERATE/SHOW** worktree directories in your
  final results/summaries because worktree directories
  are discarded after a task is completed.  Rather show
  the directories relative to the main repo.
- Authenticate unauthenticated third-party agents; ask
  the user only when a page requires human
  authentication.  You MUST collect any security or
  authentication code or token.
\end{prompt}

These overrides serve eight purposes.  First, they specify the exact toolchain commands for the KISS project itself (\texttt{uv run check -{}-full}, \texttt{uv run pytest -v} with a 900-second timeout), eliminating guesswork about which linter, formatter, or test runner to use.  Second, the agent is told the on-disk location of its own \texttt{SYSTEM.md} (under the bundled VS~Code extension directory, ending in \texttt{/SYSTEM.md}) so that questions about its own prompt or behavior can be answered by reading the canonical file rather than from memory.  This pointer is paired with a URL to the paper's \LaTeX{} source, giving the agent two complementary self-references: the system prompt for operational rules and the paper for design rationale.  Third, the instructions expose a third-party agent integration layer: the agent is told where third-party agents live (\texttt{kiss/agents/third\_party\_agents}).  When a third-party agent requires authentication, the agent handles it autonomously and only prompts the user when a page genuinely requires human credentials; the explicit rider ``You MUST collect any security or authentication code or token'' instructs the agent to elicit and persist any code/token returned during such an interactive auth step so that subsequent calls do not have to re-prompt the user.  Fourth, Claude SKILLS~\citep{claudecode2025} are made discoverable at \texttt{kiss/agents/claude\_skills} (populated at install time from the bundled VS~Code extension) and at the standard Anthropic locations \texttt{\textasciitilde/.claude/skills} and \texttt{<project>/.claude/skills}, with the note ``You can use them as necessary,'' giving the agent access to Anthropic's skill library for common software engineering patterns; the permissive framing (rather than a mandate) lets the agent select skills opportunistically when a task matches one.  Fifth, the \texttt{SORCAR.md} override mechanism (enforced by the Mandatory First Actions section) is the per-repository override file that, when present and non-empty, can extend or override the general system prompt, forming a hierarchy: general system prompt $\rightarrow$ Sorcar-specific instructions $\rightarrow$ repository-specific \texttt{SORCAR.md}.  We deliberately ship the canonical repository's \texttt{SORCAR.md} as an empty placeholder so that no override content is hardcoded inside the framework; each user repository is free to populate it with whatever project-specific instructions are appropriate, and may further \texttt{@include} additional markdown files from \texttt{SORCAR.md} for richer per-repository documentation.  Sixth, a directory-with-git-commit rule requires that any artifact the user can use after the task ends be created inside a directory and added to git, ensuring that durable outputs are version-controlled and easy to locate rather than scattered as loose files.  Seventh, the agent is required to maintain a \texttt{./tmp/PROGRESS.md} across sub-sessions logging every step taken so far with explanations and relevant code snippets; this acts as a persistent scratchpad that bridges Relentless Agent continuations and lets a fresh sub-session pick up exactly where the previous one left off without re-deriving prior decisions.  Eighth, the worktree-presentation rule (``DO NOT GENERATE/SHOW worktree directories in your final results/summaries... rather show the directories relative to the main repo'') prevents the agent from quoting transient worktree paths that are discarded after merge; user-visible references must point at paths in the main repository so that copy-pasted file names and shell commands remain valid after the task completes.

%----------------------------------------------------------------------
\section{Painless Software Engineering with KISS Sorcar}
\label{sec:painless}
%----------------------------------------------------------------------

A central claim of our system is that natural-language interaction can replace manual code inspection and ad hoc scripting for understanding and evolving nontrivial subsystems.  While developing KISS Sorcar, we found two recurring workflows particularly useful: \emph{conversational redesign}, in which the developer asks the agent to explain a subsystem and then directs incremental changes in plain English; and \emph{invariant-driven bug discovery and fixing}, in which the developer states a user-visible invariant the system MUST satisfy and asks the agent to reproduce any violation with an end-to-end test before fixing it. We illustrate both with real sessions drawn from the project's own \texttt{sorcar.db} task history, reproducing the prompts verbatim and summarizing the agent's responses.

\subsection{Conversational Redesign via Natural Language}
\label{sec:painless-redesign}

In this workflow we first ask KISS Sorcar to generate a detailed, step-by-step description of a workflow or algorithm we found buggy, and then we ask KISS Sorcar to revise some of the buggy steps in natural language. We illustrate this with a real development session in which the worktree merge workflow (Section~\ref{sec:worktree-agent}) was first understood and then redesigned entirely through conversational prompts.  The session comprises four consecutive tasks; we reproduce the prompts verbatim and summarize the agent's responses.

\paragraph{Step~1: Understanding the existing workflow.} The developer begins by asking the agent to explain the current post-task git lifecycle:

\begin{prompt}
Can you tell me what happens, step by step, with git in worktree_sorcar_agent.py when a task finishes?
\end{prompt}

\noindent The agent reads the source code and returns a structured summary of the four-phase lifecycle: (1)~during \texttt{run()}, a new branch and worktree are created and the task executes inside the worktree; when the task completes, \emph{nothing is committed or merged}; the result is returned with merge instructions appended and the worktree stays pending; (2)~\texttt{merge()} calls \texttt{\_finalize\_worktree()}, which stages all changes, generates a commit message via the LLM, commits, removes the worktree, checks out the original branch, and runs \texttt{git merge}; (3)~\texttt{discard()} removes the worktree, prunes, checks out the original branch, and deletes the task branch; (4)~in CLI mode an interactive prompt forces the user to choose \texttt{[c]ommit and merge} or \texttt{[d]iscard} before exiting. The agent also notes a key design invariant: nothing auto-merges; auto-commit occurs only at merge/finalize time; all steps are idempotent; and state can be recovered from git on restart.

\paragraph{Step~2: Simplifying the workflow via natural language.} Armed with the workflow description, the developer decides the three-way choice (auto-merge, manual merge, discard) is unnecessarily complex and issues a redesign request:

\begin{prompt}
Can you change worktree_sorcar_agent.py and the extension so that after the agent 
finishes its task, it simply asks "Commit and Merge" or "Discard". When "Commit 
and Merge" is clicked by the user, the agent must commit the changes with a 
generated commit message, merge the branch with the original branch, and delete 
the worktree and the branch associated with the worktree. If the user clicks 
"discard", it must delete the worktree and the branch, and checkout the 
original branch.
\end{prompt}

\noindent The agent modifies six files across Python and TypeScript: it updates \texttt{discard()} to check out the original branch before deleting the task branch, removes the \texttt{manual\_merge()} method entirely, simplifies \texttt{merge\_instructions()} to show only two options, updates the webview UI to replace the three-button toolbar with a two-button ``Commit and Merge or Discard?'' bar, removes the \texttt{manual} action type from the TypeScript type definitions, and removes the corresponding handler from the Python backend.  Three tests for the deleted manual-merge path have been removed, and one routing test has been updated. All 28 worktree tests pass after the change.

\paragraph{Step~3: Investigating unexpected post-merge state.} After testing the simplified flow, the developer notices that files appear in the Source Control panel of VS~Code on the original branch after pressing ``Commit and Merge'' and asks:

\begin{prompt}
After the user presses "Commit and Merge", why do the modified files show up as 
committed in the original branch?
\end{prompt}

\noindent The agent traces the exact execution path through \texttt{squash\_merge\_branch()} and discovers that the implementation deliberately \emph{unstages} the squash-merged changes via \texttt{git reset HEAD} so they appear as uncommitted working-tree modifications in VS~Code's Source Control panel.  The design intention was to let the user review the diff before committing manually.  The agent reports this finding along with the relevant code snippet and a verified end state showing no new commits on the branch.

\paragraph{Step~4: Directing a design change in one sentence.} The developer realizes that code review already happened while the worktree was active and decides the unstage step is unnecessary:

\begin{prompt}
The review is already happening in the worktree branch. You don't need the user 
to review the modified and new files in the original branch. Fix it.
\end{prompt}

\noindent The agent replaces the \texttt{git reset HEAD} call in \texttt{squash\_merge\_branch()} with a conditional \texttt{git commit --no-edit} that uses the auto-generated squash message.  It adds a \texttt{git diff --cached --quiet} guard for the edge case where the merge produces no changes.  One existing test (\texttt{test\_merge\_leaves\_changes\_uncommitted}) is renamed to \texttt{test\_merge\_commits\_changes} and its assertions are updated to verify a clean working tree.  All 104 worktree tests pass.

\medskip
\noindent\textbf{Discussion.} The four-step session illustrates the development loop that our system supports.  The developer never opens a source file, never writes a line of code, and never runs a test manually.  Instead, the entire cycle (understand the workflow, redesign it, investigate an anomaly, direct a fix) happens through natural-language prompts, with the agent handling code reading, multi-file editing, test updates, and verification.  This style of development becomes possible because of the agent hierarchy described in Section~\ref{sec:architecture}: the Worktree Sorcar Agent isolates changes on a branch, the Chat Sorcar Agent preserves conversational context across tasks, and the Relentless Agent automatically continues when the context window is exhausted.

\subsection{Invariant-Driven Bug Discovery and Fixing}
\label{sec:painless-invariants}

A second workflow we use frequently is \emph{invariant-driven bug discovery and fixing}.  The developer never localizes the defect or even names the suspect file.  Instead, the prompt (i)~states a user-visible invariant that the system MUST satisfy, (ii)~asks the agent to reproduce any violation by writing an end-to-end test before touching the source, (iii)~asks for the fix, and (iv)~assigns one model to write the test and the fix and a second model from a different vendor to review and debug that work, explicitly asking the reviewer to look for missed call sites and newly introduced bugs.  Phrasing the bug as an invariant transfers diagnostic effort to the agent; requiring an executable reproduction before the fix produces an artifact the reviewer model can refute or confirm independently of the test author's narrative; and the cross-vendor split catches mistakes that a single model would systematically miss.

We illustrate the pattern with a recent task drawn verbatim from \texttt{sorcar.db}, in which two invariants of the interactive \texttt{sorcar} CLI's input box were stated together:

\begin{prompt}
in sorcar cli interactive, the user MUST be able to enter multi-line inputs in
the input box.  Moreover, all text shown must be word wrapped.  Reproduce the
issue by writing an integration test. Then fix the issue. Use claude-opus-4-7
model for all tasks including coding, bug fixing, and test creation. Use gpt-5.5
model (not codex) for thorough review and debugging of the work done by the
other model. Check if the other model has missed some code or has introduced
bugs.
\end{prompt}

\noindent The agent executes the workflow in three phases.

\paragraph{Phase~1: Reproduce.} Under \texttt{claude-opus-4-7} the agent locates the prompt loop in \texttt{src/kiss/agents/sorcar/cli\_prompt.py}, writes a new end-to-end test file \texttt{test\_cli\_multiline\_input.py} containing nine integration tests that drive a real \texttt{prompt\_toolkit} \texttt{PromptSession} through \texttt{create\_pipe\_input} (the same pattern as the existing \texttt{test\_at\_mention\_picker.py}), and confirms the first test fails: typing \texttt{"hello"}, then Esc+Enter, then \texttt{"world"}, then Enter returns just \texttt{"hello"} against the unmodified code, witnessing the multi-line invariant violation.  The diagnosis is that \texttt{PromptSession} was instantiated without \texttt{multiline=True}, so Enter always submitted and the buffer never wrapped.

\paragraph{Phase~2: Fix.} Still under \texttt{claude-opus-4-7}, the agent passes \texttt{multiline=True}, \texttt{wrap\_lines=True}, and a \texttt{prompt\_continuation} callable to the session; the continuation callable returns an ANSI string that repaints the cyan \texttt{\textbar} left border on every wrapped or continuation row so the framed panel stays visually consistent.  An Enter binding filtered by \texttt{\textasciitilde completion\_is\_selected} calls \texttt{event.current\_buffer.validate\_and\_handle()} so plain Enter still submits, while \texttt{escape,enter} (Alt+Enter), \texttt{c-j} (Ctrl+J), and the kitty/foot/WezTerm CSI-u sequence \texttt{ESC[13;2u} (matched as a raw-char tuple because \texttt{prompt\_toolkit} does not pre-map it) insert real newlines.  The welcome banner in \texttt{cli\_repl.py} is updated to document the new keys.

\paragraph{Phase~3: Cross-model review.} The agent switches to \texttt{gpt-5.5} (not Codex) and re-reads the diff, the surrounding modules, and the new tests, looking specifically for missed call sites and broken edge cases.  The review surfaces a real \texttt{prompt\_toolkit} library limitation that \texttt{claude-opus-4-7} had failed to notice: the xterm \texttt{modifyOtherKeys} Shift+Enter escape \texttt{\textbackslash x1b[27;2;13\textasciitilde} is pre-mapped to \texttt{Keys.ControlM} inside \texttt{prompt\_toolkit.input.ansi\_escape\_sequences.ANSI\_SEQUENCES}, which means any user binding for it can never fire.  A previously-added (always-failing) test asserting that this sequence inserts a newline is replaced with one asserting the documented fall-through behaviour (it submits like plain Enter), and the welcome banner is amended to direct users on those terminals to Alt+Enter or Ctrl+J.  The review also corrects ruff E402/I001 import-ordering violations introduced during the fix.  Verification runs \texttt{uv run pytest -v} on the new file (9/9 pass) and on the neighbouring CLI suites \texttt{test\_at\_mention\_picker.py}, \texttt{test\_cli\_repl.py}, and \texttt{test\_cli\_panel.py} (49/49 pass), followed by \texttt{uv run check --full} (ruff, mypy, mdformat, syntax, api-docs) before the agent reports completion.

\medskip
\noindent\textbf{Discussion.} The pattern generalizes.  Recent entries in \texttt{sorcar.db} apply the same five-line template (``\emph{X MUST hold. Reproduce by an end-to-end test. Then fix. Use model~A for code, model~B for review.}'') to fix invariant violations in the Sorcar CLI's notification stream, in the routing of bash tool outputs to the Result panel, and in syntax-highlighting the output of the \texttt{Read} tool.  Three properties make the loop effective.  First, by phrasing the bug as an invariant rather than as a stack trace, the developer transfers diagnostic effort to the agent without having to localize the defect in advance.  Second, requiring an executable end-to-end test \emph{before} the fix produces a reproducible artifact that the reviewer model can refute or confirm independently of the implementer's narrative, and that survives in the test suite as a regression guard.  Third, splitting code production and code review across two vendors with different training data catches mistakes that a single model would systematically miss, including library-internal assumptions, such as the \texttt{prompt\_toolkit} pre-mapping above, that no amount of self-reflection by one model is likely to surface.  Dynamic model switching (Section~\ref{sec:vscode-features}) makes both halves of this split a single \texttt{set\_model()} call inside one continuous task.

%----------------------------------------------------------------------
\section{Related Work}
\label{sec:related}
%----------------------------------------------------------------------

\textbf{Code-specialized language models.} Beyond the general-purpose LLMs that our system can use as backends, a rich line of work has produced models specifically trained for code. Code Llama~\citep{roziere2023code} fine-tunes Llama~2 for code generation and infilling. StarCoder~\citep{li2023starcoder} trains on permissively licensed code from GitHub with a fill-in-the-middle objective. DeepSeek-Coder~\citep{guo2024deepseek} trains on a 2-trillion-token corpus of code and natural language with a repository-level context window. More recently, frontier models have been optimized specifically for agentic software engineering. Claude Opus~4.6~\citep{anthropic2026opus46} and its successor Opus~4.7~\citep{anthropic2026opus47} advance long-horizon coding and agentic task execution; OpenAI's GPT~5.5~\citep{openai2026gpt55} similarly targets agentic software engineering with strong code generation and tool-use capabilities. Cursor released Composer~2~\citep{cursor2026composer2}, a custom fine-tuned coding model trained with large-scale reinforcement learning. Kimi K2.5~\citep{kimiteam2026k25} is an open-source multimodal agentic model that jointly optimizes text and vision and introduces Agent Swarm for parallel task decomposition. GLM-5.1~\citep{zai2026glm51} is a 754-billion-parameter mixture-of-experts model from Z.ai that sustains autonomous coding over multi-hour sessions, achieving state-of-the-art on SWE-Bench~Pro. Our system is model-agnostic and can leverage any of these models through its pluggable LLM backend, benefiting from advances in code-specialized pre-training without architectural changes.

\textbf{Code generation agents.} SWE-Agent~\citep{yang2024sweagent} and OpenHands~\citep{wang2024openhands} provide LLM-based agents for resolving software engineering tasks such as GitHub issues.  Both use a single-session execution model without automatic continuation. Agentless~\citep{xia2024agentless} takes the opposite approach, showing that a simple two-phase localize-then-repair pipeline without autonomous agent loops can achieve competitive results on SWE-bench~\citep{jimenez2024swebench}.  Devin~\citep{devin2024}, an industrial product marketed as ``the first AI software engineer,'' operates in a sandboxed environment with shell, browser, and editor access.  Claude Code~\citep{claudecode2025} is Anthropic's terminal-based agentic coding tool that gives Claude direct access to a shell, file system, and development tools for multi-step engineering tasks.  OpenAI's Codex~\citep{codex2025} is a cloud-based software engineering agent that executes coding tasks in sandboxed environments with full repository context.  Aider~\citep{aider2023} provides a terminal-based pair programming interface with tight git integration, automatically committing each change.  Our Relentless Agent layer addresses the single-session limitation common to most of these systems, while our worktree isolation provides stronger safety guarantees than per-change commits do.

\textbf{ReAct and tool use.} The ReAct framework~\citep{yao2023react} interleaves reasoning and action. Toolformer~\citep{schick2023toolformer} teaches models to use tools via self-supervised learning.  We use native function calling provided by modern LLM APIs rather than in-context tool descriptions, thereby reducing prompt overhead and improving reliability.

\textbf{Reasoning and planning.} Chain-of-thought prompting~\citep{wei2022chain} demonstrated that eliciting step-by-step reasoning dramatically improves LLM performance on complex tasks.  Tree of Thoughts~\citep{yao2023tree} generalizes this to deliberate search over multiple reasoning paths. Reflexion~\citep{shinn2023reflexion} introduces verbal reinforcement learning, where an agent reflects on failed attempts and produces self-critiques that improve subsequent trials.  Our continuation protocol is conceptually related to Reflexion: failed sub-sessions produce summaries that inform subsequent attempts. However, in our case, we use the summaries to continue the task.

\textbf{Multi-agent software development.} ChatDev~\citep{qian2024chatdev} models the software development process as a conversation between role-playing agents (CEO, CTO, programmer, tester) organized in a waterfall-like pipeline. MetaGPT~\citep{hong2024metagpt} takes a meta-programming approach, encoding standard operating procedures as structured outputs that coordinate specialized agents. AutoGen~\citep{wu2023autogen} provides a general-purpose framework for multi-agent conversation, enabling flexible topologies beyond fixed pipelines.  These systems focus on generating entire applications from scratch.  We take a different approach: rather than simulating an organization of specialists, we use a single agent with broad access to tools and optional parallel sub-agents for embarrassingly parallel sub-tasks, prioritizing practical utility on real-world codebases over role-playing fidelity.

\textbf{Multi-turn autonomous agents.} AutoGPT~\citep{autogpt2023} and BabyAGI~\citep{babyagi2023} implement multi-step autonomous agents.  These systems typically lack budget controls and safe code isolation.  Our layered architecture addresses each of these concerns in a dedicated layer.

\textbf{Software engineering benchmarks.} SWE-bench~\citep{jimenez2024swebench} evaluates agents on real-world GitHub issues drawn from popular Python repositories, requiring the agent to localize and fix bugs given only the issue description. It has become the de facto standard for measuring agent capabilities in software engineering. HumanEval~\citep{chen2021evaluating} and MBPP~\citep{austin2021program} evaluate function-level code generation from docstrings. LiveCodeBench~\citep{jain2024livecodebench} addresses benchmark contamination by continuously collecting fresh competition problems from LeetCode, AtCoder, and CodeForces, and extends evaluation to self-repair, code execution, and test output prediction.  These benchmarks focus on isolated coding problems; We target the broader workflow of multi-file, multi-step software engineering tasks that require tool use, testing, and version control.

\textbf{Agent frameworks and orchestration.} LangChain~\citep{langchain2022} provides modular abstractions for building LLM applications, including agent loops, tool integration, and memory. It focuses on composability and breadth of integrations rather than on the specific concerns of software development.  DSPy~\citep{khattab2024dspy} treats LLM calls as declarative modules whose prompts and few-shot examples are compiled and optimized automatically, enabling systematic improvement of multi-stage pipelines; it targets prompt optimization rather than end-to-end software engineering workflows.  CrewAI~\citep{crewai2024} orchestrates role-playing autonomous agents that collaborate through configurable process models and event-driven flows, emphasizing multi-agent coordination over the layered safety and budget controls that we prioritize.  smolagents~\citep{smolagents2025}, Hugging Face's minimalist agent library, has its CodeAgent write actions as executable Python snippets, reducing step counts by~30\%; however, it does not address repository-level concerns such as git isolation or continuation across sessions.  The OpenAI Agents SDK~\citep{openaiagentssdk2025} offers a provider-agnostic framework for multi-agent workflows with handoffs, guardrails, and tracing, while Google's Agent Development Kit~\citep{googleadk2025} provides an open-source toolkit for building, evaluating, and deploying AI agents with multi-agent orchestration.  Both provide general-purpose agent infrastructure but leave domain-specific concerns to the application layer.  Our architecture is purpose-built for software engineering and general research, with each layer addressing a specific practical concern (budget tracking, continuation, code safety).

\textbf{LLM agent surveys.} Several comprehensive surveys have mapped the rapidly growing landscape of LLM-based agents. \citet{xi2023rise} surveys the design space of LLM agents along three dimensions, brain (reasoning), perception (input modalities), and action (tool use), and catalogs applications across social science, natural science, and engineering. \citet{wang2024survey} propose a systematic framework for autonomous agents built on LLMs, covering profile, memory, planning, and action modules.  Our system can be understood through the lens of these frameworks: the KISS Agent implements the action loop; the Relentless Agent addresses memory and planning concerns through continuation summaries; and the system prompt encodes the profile.

\textbf{IDE integration.} GitHub Copilot~\citep{copilot2021}, Cursor~\citep{cursor2024}, and Windsurf~\citep{windsurf2024} provide AI assistance within editors. Cursor recently released Composer~2~\citep{cursor2026composer2}, a custom fine-tuned coding model trained with large-scale reinforcement learning that achieves 61.7\% on Terminal Bench~2.0. We operate at the level of autonomous multi-step task execution with full tool access and git-level isolation.

\subsection{Dynamic Steering of Running Agents}
\label{sec:dynamic-steering}

A practical autonomous assistant must remain steerable after a task has started.  KISS Sorcar therefore supports \emph{dynamic steering}: while any agent is running, the user can add a new natural-language message to the live task, and that message is delivered to the selected agent as additional user guidance.  The agent does not need to finish its current high-level task, reach an explicit permission prompt, or be restarted from a new initial prompt.  The mechanism is intentionally simple: a running task remains a conversation, and the user's later message becomes part of the task trajectory that the agent must incorporate when it next reaches a safe decision point.  This is useful when the user observes that the agent has misunderstood a requirement, discovers a new constraint, notices that an external command is taking too long, or wants to redirect a parallel exploration toward a more promising branch.

This feature is orthogonal to the continuation mechanism in Section~\ref{sec:relentless-agent}.  Relentless continuation preserves progress across context-window or step-budget boundaries; dynamic steering lets the human change the objective while progress is still being made.  It is also orthogonal to worktree isolation: steering can redirect the behavior of a task without sacrificing the ability to later commit-and-merge or discard the whole branch.  Because KISS Sorcar treats external systems such as Claude Code CLI and Codex CLI as model backends and exposes third-party messaging agents through the same tool layer, dynamic steering is not tied to a single first-party agent.  The same user action can steer a local KISS Agent, a worktree-isolated Sorcar task, or an agent that delegates work through a supported external backend.

Several recent systems provide related forms of in-progress control.  Claude Code's Remote Control allows a user to continue a local session from a browser or phone, keep the conversation synchronized across devices, and send messages interchangeably from terminal, web, and mobile surfaces~\citep{anthropic2026remotecontrol}.  This is the closest product analogue: it explicitly targets steering in-progress work from another device.  KISS Sorcar differs in scope: dynamic steering is an IDE/runtime primitive for any KISS-managed agent, including third-party backends and parallel sub-agents, rather than a remote surface for one vendor's agent.  GitHub Copilot's cloud agent can be monitored and steered with follow-up prompts in GitHub, and users can mention \texttt{@copilot} on a pull request to request further changes~\citep{github2026copilotcloud}.  However, issue comments added after initial assignment are not observed until the workflow moves to the pull request, and follow-ups often create or continue PR-bound sessions.  KISS Sorcar instead treats steering as a live message to the active task, independent of the GitHub issue/PR lifecycle.

Cursor Cloud Agents and OpenAI Codex both expose richer background-agent workflows.  Cursor can launch cloud agents from web, desktop, Slack, GitHub/Bitbucket comments, Linear, and an API, and it provides artifacts plus remote desktop control of the agent's environment~\citep{cursor2026cloudagents}.  Codex has added mobile remote workflows, queue-versus-steer follow-up behavior, side chats, and progress visibility for running threads and subagents~\citep{openai2026codexchangelog}.  These systems demonstrate that users need to follow and redirect long-running agent work.  KISS Sorcar's contribution is to make that capability small, uniform, and backend-agnostic: the user's steering message is routed through the same local task machinery as the original instruction, rather than through a product-specific cloud session, remote desktop handoff, or PR comment loop.

Other related mechanisms are more specialized.  Aider's \texttt{--watch-files} mode lets users place \texttt{AI!} or \texttt{AI?} comments in source files, which Aider collects as coding instructions from the editor~\citep{aider2026watch}.  This provides in-context steering through file edits, but it is specific to Aider and to source-code comments.  LangChain/LangGraph's human-in-the-loop middleware pauses execution around configured tool calls and resumes after a human approves, edits, rejects, or responds to the pending action~\citep{langchain2026hitl}.  Semantic Kernel similarly describes the agent loop as repeated function calling until completion or until the model needs user help~\citep{microsoft2025semantickernelplanning}.  These frameworks provide useful interrupt/resume substrates, but they center on tool-call approval or model-requested help.  Dynamic steering in KISS Sorcar is broader: the user may inject arbitrary revised task guidance even when the agent did not explicitly ask for input, allowing human intent to remain in the loop throughout long-running autonomous work.

\textbf{Recent advances in agentic software engineering.} The field has matured rapidly since late 2025. \citet{hassan2025agentic} articulate the foundational pillars of \emph{Agentic Software Engineering} (SE~3.0), defining a research roadmap that emphasizes trust, controllability, and goal-oriented task decomposition.  \citet{li2025rise} surveys the landscape of autonomous coding agents and characterizes the transition from assistive code completion to full-fledged AI teammates that initiate, plan, and execute development tasks.  We instantiate several of the principles advocated in these roadmaps, including layered controllability (via budget tracking and step limits) and safe isolation (via worktrees).

\citet{wang2025agentic} provides a comprehensive survey of AI agentic programming techniques, cataloging how LLM-based agents decompose goals, interact with compilers and version control systems, and self-correct through feedback loops. \citet{chatlatanagulchai2025readmes} study \emph{agent context files} (persistent, project-level instruction files that guide agentic coding tools) and find that high-quality context files significantly improve agent performance.  Our layered system prompt and \texttt{SORCAR.md} override mechanisms are instances of this pattern. \citet{mohsenimofidi2025context} further investigates context engineering for AI agents in open-source software, highlighting how curated context improves agent efficacy on repository-level tasks.

\textbf{Evolving benchmarks for coding agents.} SWE-bench Pro~\citep{deng2025swebenchpro} introduces a substantially more challenging benchmark with 1,865 long-horizon problems drawn from 41 repositories, including commercial codebases, explicitly targeting enterprise-level complexity beyond the original SWE-bench.  These long-horizon tasks (often requiring multi-file patches and hours of professional developer effort) directly motivate our continuation mechanism.  \citet{prathifkumar2025swebenchverified} raises the important question of benchmark contamination, showing that overlap between SWE-Bench-Verified problems and LLM training data may inflate scores, suggesting that high benchmark numbers may partly reflect memorization rather than genuine problem-solving ability. \citet{horikawa2025agentic} provides an empirical study of how AI coding agents handle refactoring tasks, finding that while agents can plan and execute complex refactorings, they still struggle with cross-file dependency analysis, a challenge that our sub-agent parallelism partially addresses.

\textbf{Self-improvement and test-time scaling.} \citet{robeyns2025selfimproving} demonstrates a self-improving coding agent that iteratively refines its own scaffolding code, achieving dramatic benchmark improvements through self-generated optimizations.  We previously experimented with a much lighter-weight variant of this idea via a per-project preferences file that the agent read and updated each session, but removed the mechanism (Section~\ref{sec:self-improvement-prompt}) because the store retained stale project facts that went out of date as the code evolved, leading the agent to reintroduce already-fixed bugs.  \citet{gao2025trae} introduces the Trae Agent, which applies test-time compute scaling to software engineering, dynamically allocating more inference budget to harder problems.  Our budget-tracking mechanism provides a complementary perspective: rather than scaling compute adaptively, we enforce hard budget ceilings while the Relentless Agent ensures maximum progress within those bounds.

\textbf{Community-driven prompt engineering.} Get Shit Done (GSD)~\citep{gsd2025} is a light-weight meta-prompting, context engineering, and spec-driven development system originally created for Claude Code and later extended to other AI coding agents.  GSD addresses \emph{context rot} (the quality degradation that occurs as an LLM fills its context window) by structuring work into phased planning documents and spawning specialized parallel agents for research, planning, execution, and verification.  Its core philosophy rejects enterprise ceremony in favor of directness: ``no enterprise roleplay bullshit,'' as the author puts it.  The sections ``Pre-flight Checks'', ``Deep Work'', and ``Pre-Finish Verification'' of \texttt{SYSTEM.md} were partly inspired by this work.

%----------------------------------------------------------------------
\section{Conclusion}
\label{sec:conclusion}
%----------------------------------------------------------------------

We have shown that a simple layered, single-concern architecture can address the practical challenges of deploying LLM agents for real-world software development.  On Terminal Bench~2.0, a benchmark of 89 diverse terminal-based tasks evaluated across 5 trials each, our system achieves a 62.2\% overall pass rate (277/445), compared with Claude~Code (58\%) and Cursor Composer~2 (61.7\%) on the same benchmark with the same underlying model (Claude Opus~4.6).  These results are achieved without benchmark-specific optimizations, fine-tuning, or reinforcement learning.

We complement the architecture with a system prompt that encodes engineering practices directly into the agent's behavior: read before writing, test before fixing, plan before executing, verify before finishing.  These are not novel insights; they are the practices of careful software engineering, translated into instructions that an LLM can follow.  The evaluation suggests that giving a frontier model the time and tools to validate its own output matters more than model-level customization.

In summary, we showed that a simple agent framework, without sophisticated agent technologies such as trajectory compaction and asynchronous multi-agent orchestration, was sufficient to build KISS Sorcar.  By building KISS Sorcar using itself and matching or exceeding both Cursor and Claude Code, we found that established software engineering techniques and principles are important for building reliable agent systems.

\section*{An Important Advice} We found that researchers are publishing 
lots of papers on AI, and it is hard to keep track of them or validate their claims. We propose to use the following prompt with KISS Sorcar to identify issues in blogs, papers, and code repositories:

\begin{prompt}
Can you read <<url>>, and thoroughly and precisely check for **wrong assumptions**, **cheating**,
**irreproducibility issues**, **fraud**, **potential for cheating in evaluation**,
**AI slop**, and **security vulnerabilities**?  
Use the internet search extensively and do not believe what people say--verify it yourself.  
Do not hesitate to download the code and run it to validate the results.
For security vulnerabilities, create a POC and test it.  
Generate an HTML report in ./sorcar_reported_frauds/ and open it in the user's default browser.
Thoroughly fact check everything you claim in the report.
\end{prompt}
In the prompt, replace <<url>> with the actual link to a blog, a paper, or a code repository.

\section*{Acknowledgments}

This research is supported in part by gifts from Accenture, Amazon, AMD, Anyscale, Broadcom, Google, IBM, Intel, Intesa Sanpaolo, Lambda, Lightspeed, Mibura, NVIDIA, Samsung SDS, SAP, by the U.S. Department of Energy, Office of Science, Office of Advanced Scientific Computing Research through the X-STACK: Programming Environments for Scientific Computing program (DESC0021982,) and the Defense Advanced Research Projects Agency (DARPA) under Agreement No. HR00112590134.

We would like to thank Marius Momeu for finding critical bugs in the cost computation and in the UI usability, Yogya Mehrotra for testing and fixing bugs in the OpenClaw like third\_party\_agents in KISS Sorcar; Debabrata Dash, Yiwei Hou, Kaiyao Ke, Muxi Lyu, Anoop Mishra, Manish Shetty, Ion Stoica, Shangyin Tan, Hao Wang, and Matei Zaharia for various useful feedback.

%----------------------------------------------------------------------
% References
%----------------------------------------------------------------------

{ \small

\bibliographystyle{plainnat}
\bibliography{kiss_sorcar}

@article{chen2021evaluating,
  author    = {Chen, Mark and Tworek, Jerry and Jun, Heewoo and Yuan, Qiming and Pinto, Henrique Pond\'{e} and Kaplan, Jared and Edwards, Harri and Burda, Yuri and Joseph, Nicholas and Brockman, Greg and others},
  title     = {Evaluating Large Language Models Trained on Code},
  journal   = {arXiv preprint arXiv:2107.03374},
  year      = {2021},
}

@article{roziere2023code,
  author    = {Rozi\`{e}re, Baptiste and Gehring, Jonas and Gloeckle, Fabian and Sootla, Sten and Gat, Itai and Tan, Xiaoqing Ellen and Adi, Yossi and Liu, Jingyu and Remez, Tal and Rapin, J\'{e}r\'{e}my and others},
  title     = {{Code Llama}: Open Foundation Models for Code},
  journal   = {arXiv preprint arXiv:2308.12950},
  year      = {2023},
}

@inproceedings{yao2023react,
  author    = {Yao, Shunyu and Zhao, Jeffrey and Yu, Dian and Du, Nan and Shafran, Izhak and Narasimhan, Karthik and Cao, Yuan},
  title     = {{ReAct}: Synergizing Reasoning and Acting in Language Models},
  booktitle = {International Conference on Learning Representations (ICLR)},
  year      = {2023},
}

@inproceedings{schick2023toolformer,
  author    = {Schick, Timo and Dwivedi-Yu, Jane and Dess\`{i}, Roberto and Raileanu, Roberta and Lomeli, Maria and Zettlemoyer, Luke and Cancedda, Nicola and Scialom, Thomas},
  title     = {Toolformer: Language Models Can Teach Themselves to Use Tools},
  booktitle = {Advances in Neural Information Processing Systems (NeurIPS)},
  year      = {2023},
}

@article{yang2024sweagent,
  author    = {Yang, John and Jimenez, Carlos E. and Wettig, Alexander and Lieret, Kilian and Yao, Shunyu and Narasimhan, Karthik and Press, Ofir},
  title     = {{SWE-agent}: Agent-Computer Interfaces Enable Automated Software Engineering},
  journal   = {arXiv preprint arXiv:2405.15793},
  year      = {2024},
}

@article{wang2024openhands,
  author    = {Wang, Xingyao and Chen, Yangyi and Yuan, Lifan and Zhang, Yizhe and Li, Yunzhi and Peng, Hao and Ji, Heng},
  title     = {{OpenHands}: An Open Platform for {AI} Software Developers as Generalist Agents},
  journal   = {arXiv preprint arXiv:2407.16741},
  year      = {2024},
}

@misc{autogpt2023,
  author    = {{Significant Gravitas}},
  title     = {{AutoGPT}},
  howpublished = {\url{https://github.com/Significant-Gravitas/AutoGPT}},
  year      = {2023},
}

@misc{codex2025,
  author    = {{OpenAI}},
  title     = {Introducing {Codex}: A Cloud-Based Software Engineering Agent},
  howpublished = {\url{https://openai.com/index/introducing-codex/}},
  year      = {2025},
}

@misc{openclaw2025,
  author    = {{OpenClaw AI}},
  title     = {{OpenClaw}: Personal {AI} Assistant},
  howpublished = {\url{https://openclaw.ai}},
  year      = {2025},
}

@misc{babyagi2023,
  author    = {Nakajima, Yohei},
  title     = {{BabyAGI}},
  howpublished = {\url{https://github.com/yoheinakajima/babyagi}},
  year      = {2023},
}

@misc{copilot2021,
  author    = {{GitHub}},
  title     = {{GitHub Copilot}: Your {AI} Pair Programmer},
  howpublished = {\url{https://github.com/features/copilot}},
  year      = {2021},
}

@misc{cursor2024,
  author    = {{Cursor}},
  title     = {{Cursor}: The {AI}-First Code Editor},
  howpublished = {\url{https://cursor.sh}},
  year      = {2024},
}

@misc{windsurf2024,
  author    = {{Codeium}},
  title     = {{Windsurf}: The {AI}-Powered {IDE}},
  howpublished = {\url{https://windsurf.com}},
  year      = {2024},
}

@inproceedings{jimenez2024swebench,
  author    = {Jimenez, Carlos E. and Yang, John and Wettig, Alexander and Yao, Shunyu and Pei, Kexin and Press, Ofir and Narasimhan, Karthik},
  title     = {{SWE-bench}: Can Language Models Resolve Real-World {GitHub} Issues?},
  booktitle = {International Conference on Learning Representations (ICLR)},
  year      = {2024},
}

@article{xia2024agentless,
  author    = {Xia, Chunqiu Steven and Deng, Yinlin and Dunn, Soren and Zhang, Lingming},
  title     = {Agentless: Demystifying {LLM}-Based Software Engineering Agents},
  journal   = {arXiv preprint arXiv:2407.01489},
  year      = {2024},
}

@misc{claudecode2025,
  author    = {{Anthropic}},
  title     = {{Claude Code}: Anthropic's Agentic Coding System},
  howpublished = {\url{https://www.anthropic.com/product/claude-code}},
  year      = {2025},
}

@misc{devin2024,
  author    = {{Cognition Labs}},
  title     = {{Devin}: The First {AI} Software Engineer},
  howpublished = {\url{https://devin.ai}},
  year      = {2024},
}

@misc{aider2023,
  author    = {Gauthier, Paul},
  title     = {{Aider}: {AI} Pair Programming in Your Terminal},
  howpublished = {\url{https://github.com/paul-gauthier/aider}},
  year      = {2023},
}

@inproceedings{wei2022chain,
  author    = {Wei, Jason and Wang, Xuezhi and Schuurmans, Dale and Bosma, Maarten and Ichter, Brian and Xia, Fei and Chi, Ed and Le, Quoc and Zhou, Denny},
  title     = {Chain-of-Thought Prompting Elicits Reasoning in Large Language Models},
  booktitle = {Advances in Neural Information Processing Systems (NeurIPS)},
  year      = {2022},
}

@inproceedings{yao2023tree,
  author    = {Yao, Shunyu and Yu, Dian and Zhao, Jeffrey and Shafran, Izhak and Griffiths, Thomas L. and Cao, Yuan and Narasimhan, Karthik},
  title     = {Tree of Thoughts: Deliberate Problem Solving with Large Language Models},
  booktitle = {Advances in Neural Information Processing Systems (NeurIPS)},
  year      = {2023},
}

@inproceedings{shinn2023reflexion,
  author    = {Shinn, Noah and Cassano, Federico and Gopinath, Ashwin and Narasimhan, Karthik and Yao, Shunyu},
  title     = {Reflexion: Language Agents with Verbal Reinforcement Learning},
  booktitle = {Advances in Neural Information Processing Systems (NeurIPS)},
  year      = {2023},
}

@inproceedings{qian2024chatdev,
  author    = {Qian, Chen and Liu, Wei and Liu, Hongzhang and Chen, Nuo and Dang, Yufan and Li, Jiahao and Yang, Cheng and Chen, Weize and Su, Yusheng and Cong, Xin and Xu, Juyuan and Li, Dahai and Liu, Zhiyuan and Sun, Maosong},
  title     = {{ChatDev}: Communicative Agents for Software Development},
  booktitle = {Proceedings of the 62nd Annual Meeting of the Association for Computational Linguistics (ACL)},
  year      = {2024},
}

@inproceedings{hong2024metagpt,
  author    = {Hong, Sirui and Zhuge, Mingchen and Chen, Jonathan and Zheng, Xiawu and Cheng, Yuheng and Zhang, Ceyao and Wang, Jinlin and Wang, Zili and Yau, Steven Ka Shing and Lin, Zijuan and Zhou, Liyang and Ran, Chenyu and Xiao, Lingfeng and Wu, Chenglin and Schmidhuber, J\"{u}rgen},
  title     = {{MetaGPT}: Meta Programming for a Multi-Agent Collaborative Framework},
  booktitle = {International Conference on Learning Representations (ICLR)},
  year      = {2024},
}

@article{wu2023autogen,
  author    = {Wu, Qingyun and Bansal, Gagan and Zhang, Jieyu and Wu, Yiran and Li, Beibin and Zhu, Erkang and Jiang, Li and Zhang, Xiaoyun and Zhang, Shaokun and Liu, Jiale and Awadallah, Ahmed Hassan and White, Ryen W. and Burger, Doug and Wang, Chi},
  title     = {{AutoGen}: Enabling Next-Gen {LLM} Applications via Multi-Agent Conversation},
  journal   = {arXiv preprint arXiv:2308.08155},
  year      = {2023},
}

@article{li2023starcoder,
  author    = {Li, Raymond and Allal, Loubna Ben and Zi, Yangtian and Muennighoff, Niklas and Kocetkov, Denis and Mou, Chenghao and Marone, Marc and Akiki, Christopher and Li, Jia and Chim, Jenny and others},
  title     = {{StarCoder}: May the Source Be with You!},
  journal   = {Transactions on Machine Learning Research (TMLR)},
  year      = {2023},
}

@article{guo2024deepseek,
  author    = {Guo, Daya and Zhu, Qihao and Yang, Dejian and Xie, Zhenda and Dong, Kai and Zhang, Wentao and Chen, Guanting and Bi, Xiao and Wu, Y. and Li, Y. K. and Luo, Fuli and Xiong, Yun and Liang, Wenfeng},
  title     = {{DeepSeek-Coder}: When the Large Language Model Meets Programming --- The Rise of Code Intelligence},
  journal   = {arXiv preprint arXiv:2401.14196},
  year      = {2024},
}

@article{austin2021program,
  author    = {Austin, Jacob and Odena, Augustus and Nye, Maxwell and Bosma, Maarten and Michalewski, Henryk and Dohan, David and Jiang, Ellen and Cai, Carrie and Terry, Michael and Le, Quoc and Sutton, Charles},
  title     = {Program Synthesis with Large Language Models},
  journal   = {arXiv preprint arXiv:2108.07732},
  year      = {2021},
}

@misc{langchain2022,
  author    = {{LangChain}},
  title     = {{LangChain}: Build Context-Aware Reasoning Applications},
  howpublished = {\url{https://github.com/langchain-ai/langchain}},
  year      = {2022},
}

@article{xi2023rise,
  author    = {Xi, Zhiheng and Chen, Wenxiang and Guo, Xin and He, Wei and Ding, Yiwen and Hong, Boyang and Zhang, Ming and Wang, Junzhe and Jin, Senjie and Zhou, Enyu and others},
  title     = {The Rise and Potential of Large Language Model Based Agents: A Survey},
  journal   = {arXiv preprint arXiv:2309.07864},
  year      = {2023},
}

@article{wang2024survey,
  author    = {Wang, Lei and Ma, Chen and Feng, Xueyang and Zhang, Zeyu and Yang, Hao and Zhang, Jingsen and Chen, Zhiyuan and Tang, Jiakai and Chen, Xu and Lin, Yankai and Zhao, Wayne Xin and Wei, Zhewei and Wen, Ji-Rong},
  title     = {A Survey on Large Language Model Based Autonomous Agents},
  journal   = {Frontiers of Computer Science},
  volume    = {18},
  number    = {6},
  pages     = {186345},
  year      = {2024},
}

@article{hassan2025agentic,
  author    = {Hassan, Ahmed E. and Li, Hao and Lin, Dayi and Adams, Bram and Chen, Tse-Hsun and Kashiwa, Yutaro and Qiu, Dong},
  title     = {Agentic Software Engineering: Foundational Pillars and a Research Roadmap},
  journal   = {arXiv preprint arXiv:2509.06216},
  year      = {2025},
}

@article{li2025rise,
  author    = {Li, Hao and Zhang, Haoxiang and Hassan, Ahmed E.},
  title     = {The Rise of {AI} Teammates in Software Engineering ({SE}~3.0): How Autonomous Coding Agents are Reshaping Software Engineering},
  journal   = {arXiv preprint arXiv:2507.15003},
  year      = {2025},
}

@article{deng2025swebenchpro,
  author    = {Deng, Xiang and Da, Jeff and Pan, Edwin and He, Yannis Yiming and Ide, Charles and Garg, Kanak and Lauffer, Niklas and Park, Andrew and Pasari, Nitin and Rane, Chetan and others},
  title     = {{SWE-Bench Pro}: Can {AI} Agents Solve Long-Horizon Software Engineering Tasks?},
  journal   = {arXiv preprint arXiv:2509.16941},
  year      = {2025},
}

@article{prathifkumar2025swebenchverified,
  author    = {Prathifkumar, Thanosan and Mathews, Noble Saji and Nagappan, Meiyappan},
  title     = {Does {SWE-Bench-Verified} Test Agent Ability or Model Memory?},
  journal   = {arXiv preprint arXiv:2512.10218},
  year      = {2025},
}

@article{horikawa2025agentic,
  author    = {Horikawa, Kosei and Li, Hao and Kashiwa, Yutaro and Adams, Bram and Iida, Hajimu and Hassan, Ahmed E.},
  title     = {Agentic Refactoring: An Empirical Study of {AI} Coding Agents},
  journal   = {arXiv preprint arXiv:2511.04824},
  year      = {2025},
}

@article{wang2025agentic,
  author    = {Wang, Huanting and Gong, Jingzhi and Zhang, Huawei and Xu, Jie and Wang, Zheng},
  title     = {{AI} Agentic Programming: A Survey of Techniques, Challenges, and Opportunities},
  journal   = {arXiv preprint arXiv:2508.11126},
  year      = {2025},
}

@article{chatlatanagulchai2025readmes,
  author    = {Chatlatanagulchai, Worawalan and Li, Hao and Kashiwa, Yutaro and Reid, Brittany and Thonglek, Kundjanasith and Leelaprute, Pattara and Rungsawang, Arnon and Manaskasemsak, Bundit and Adams, Bram and Hassan, Ahmed E. and Iida, Hajimu},
  title     = {Agent {README}s: An Empirical Study of Context Files for Agentic Coding},
  journal   = {arXiv preprint arXiv:2511.12884},
  year      = {2025},
}

@article{mohsenimofidi2025context,
  author    = {Mohsenimofidi, Seyedmoein and Galster, Matthias and Treude, Christoph and Baltes, Sebastian},
  title     = {Context Engineering for {AI} Agents in Open-Source Software},
  journal   = {arXiv preprint arXiv:2510.21413},
  year      = {2025},
}

@article{robeyns2025selfimproving,
  author    = {Robeyns, Maxime and Szummer, Martin and Aitchison, Laurence},
  title     = {A Self-Improving Coding Agent},
  journal   = {arXiv preprint arXiv:2504.15228},
  year      = {2025},
}

@article{gao2025trae,
  author    = {Gao, Pengfei and Tian, Zhao and Meng, Xiangxin and {Trae Research Team}},
  title     = {Trae Agent: An {LLM}-Based Agent for Software Engineering with Test-Time Scaling},
  journal   = {arXiv preprint arXiv:2507.23370},
  year      = {2025},
}

@article{cursor2026composer2,
  author    = {{Cursor Research}},
  title     = {Composer~2 Technical Report},
  journal   = {arXiv preprint arXiv:2603.24477},
  year      = {2026},
}

@inproceedings{agrawal2025gepa,
  author    = {Agrawal, Lakshya A. and Tan, Shangyin and Soylu, Dilara and Ziems, Noah and Khare, Rishi and Opsahl-Ong, Krista and Singhvi, Arnav and Shandilya, Herumb and Ryan, Michael J. and Jiang, Meng and Potts, Christopher and Sen, Koushik and Dimakis, Alexandros G. and Stoica, Ion and Klein, Dan and Zaharia, Matei and Khattab, Omar},
  title     = {{GEPA}: Reflective Prompt Evolution Can Outperform Reinforcement Learning},
  booktitle = {International Conference on Learning Representations (ICLR)},
  year      = {2026},
  note      = {Oral. arXiv preprint arXiv:2507.19457},
}

@article{romera2024funsearch,
  author    = {Romera-Paredes, Bernardino and Barekatain, Mohammadamin and Novikov, Alexander and Balog, Matej and Kumar, M. Pawan and Dupont, Emilien and Ruiz, Francisco J. R. and Ellenberg, Jordan S. and Wang, Pengming and Fawzi, Omar and Kohli, Pushmeet and Fawzi, Alhussein},
  title     = {Mathematical discoveries from program search with large language models},
  journal   = {Nature},
  volume    = {625},
  pages     = {468--475},
  year      = {2024},
  doi       = {10.1038/s41586-023-06924-6},
}

@article{novikov2025alphaevolve,
  author    = {Novikov, Alexander and V\~{u}, Ng\^{a}n and Eisenberger, Marvin and Dupont, Emilien and Huang, Po-Sen and Wagner, Adam Zsolt and Shirobokov, Sergey and Kozlovskii, Borislav and Ruiz, Francisco J. R. and Mehrabian, Abbas and Kumar, M. Pawan and See, Abigail and Chaudhuri, Swarat and Holland, George and Davies, Alex and Nowozin, Sebastian and Kohli, Pushmeet and Balog, Matej},
  title     = {{AlphaEvolve}: A Coding Agent for Scientific and Algorithmic Discovery},
  journal   = {arXiv preprint arXiv:2506.13131},
  year      = {2025},
}

@misc{openevolve2025,
  author    = {{Algorithmic Superintelligence}},
  title     = {{OpenEvolve}: Open-Source Implementation of {AlphaEvolve}},
  howpublished = {\url{https://github.com/algorithmicsuperintelligence/openevolve}},
  year      = {2025},
}

@article{lange2025shinka,
  author    = {Lange, Robert Tjarko and Imajuku, Yuki and Cetin, Edoardo},
  title     = {{ShinkaEvolve}: Towards Open-Ended and Sample-Efficient Program Evolution},
  journal   = {arXiv preprint arXiv:2509.19349},
  year      = {2025},
}

@article{cheng2025adrs,
  author    = {Cheng, Audrey and Liu, Shu and Pan, Melissa and Li, Zhifei and Wang, Bowen and Krentsel, Alex and Xia, Tian and Cemri, Mert and Park, Jongseok and Yang, Shuo and Chen, Jeff and Agrawal, Lakshya and Desai, Aditya and Xing, Jiarong and Sen, Koushik and Zaharia, Matei and Stoica, Ion},
  title     = {Barbarians at the Gate: How {AI} is Upending Systems Research},
  journal   = {arXiv preprint arXiv:2510.06189},
  year      = {2025},
}

@misc{stein2026cheatingblog,
  author    = {Stein, Adam and Brown, Davis and Hassani, Hamed and Naik, Mayur and Wong, Eric},
  title     = {Finding Widespread Cheating on Popular Agent Benchmarks},
  howpublished = {Blog post, \url{https://debugml.github.io/cheating-agents/}},
  year      = {2026},
}

@article{stein2026detecting,
  author    = {Stein, Adam and Brown, Davis and Hassani, Hamed and Naik, Mayur and Wong, Eric},
  title     = {Detecting Safety Violations Across Many Agent Traces},
  journal   = {arXiv preprint arXiv:2604.11806},
  year      = {2026},
}

@misc{wang2026trustworthyblog,
  author    = {Wang, Hao and Mang, Qiuyang and Cheung, Alvin and Sen, Koushik and Song, Dawn},
  title     = {We Scored 100\% on {AI} Benchmarks Without Solving a Single Problem},
  howpublished = {Blog post, \url{https://moogician.github.io/blog/2026/trustworthy-benchmarks/}},
  year      = {2026},
}

@article{kimiteam2026k25,
  author    = {{Kimi Team}},
  title     = {Kimi {K2.5}: Visual Agentic Intelligence},
  journal   = {arXiv preprint arXiv:2602.02276},
  year      = {2026},
}

@misc{zai2026glm51,
  author    = {{Z.ai}},
  title     = {{GLM-5.1}: Towards Long-Horizon Tasks},
  howpublished = {Technical blog, \url{https://z.ai/blog/glm-5.1}},
  year      = {2026},
}

@misc{anthropic2026opus46,
  author    = {{Anthropic}},
  title     = {Introducing {Claude} {Opus}~4.6},
  howpublished = {\url{https://www.anthropic.com/news/claude-opus-4-6}},
  year      = {2026},
}

@misc{anthropic2026opus47,
  author    = {{Anthropic}},
  title     = {Introducing {Claude} {Opus}~4.7},
  howpublished = {\url{https://www.anthropic.com/news/claude-opus-4-7}},
  year      = {2026},
}

@misc{openai2026gpt55,
  author    = {{OpenAI}},
  title     = {Introducing {GPT}-5.5},
  howpublished = {\url{https://openai.com/index/introducing-gpt-5-5/}},
  year      = {2026},
}

@misc{google2026gemini31,
  author    = {{Google DeepMind}},
  title     = {{Gemini}~3.1 {Pro}},
  howpublished = {\url{https://deepmind.google/models/gemini/pro/}},
  year      = {2026},
}

@article{jain2024livecodebench,
  author    = {Naman Jain and King Han and Alex Gu and Wen-Ding Li and Fanjia Yan and Tianjun Zhang and Sida Wang and Armando Solar-Lezama and Koushik Sen and Ion Stoica},
  title     = {{LiveCodeBench}: Holistic and Contamination Free Evaluation of Large Language Models for Code},
  journal   = {arXiv preprint arXiv:2403.07974},
  year      = {2024},
}

@inproceedings{khattab2024dspy,
  title     = {{DSPy}: Compiling Declarative Language Model Calls into Self-Improving Pipelines},
  author    = {Khattab, Omar and Singhvi, Arnav and Maheshwari, Paridhi and Zhang, Zhiyuan and Santhanam, Keshav and Vardhamanan, Sri and Haq, Saiful and Sharma, Ashutosh and Joshi, Thomas T. and Moazam, Hanna and Miller, Heather and Zaharia, Matei and Potts, Christopher},
  booktitle = {The Twelfth International Conference on Learning Representations},
  year      = {2024},
}

@misc{crewai2024,
  author       = {{CrewAI, Inc.}},
  title        = {{CrewAI}: Framework for Orchestrating Role-Playing, Autonomous {AI} Agents},
  howpublished = {\url{https://github.com/crewAIInc/crewAI}},
  year         = {2024},
}

@misc{smolagents2025,
  author       = {Roucher, Aymeric and Villanova del Moral, Albert and Wolf, Thomas and von Werra, Leandro and Kaunism{\"a}ki, Erik},
  title        = {smolagents: A Smol Library to Build Great Agentic Systems},
  howpublished = {\url{https://github.com/huggingface/smolagents}},
  year         = {2025},
}

@misc{openaiagentssdk2025,
  author       = {{OpenAI}},
  title        = {{OpenAI Agents SDK}: A Lightweight, Powerful Framework for Multi-Agent Workflows},
  howpublished = {\url{https://github.com/openai/openai-agents-python}},
  year         = {2025},
}

@misc{googleadk2025,
  author       = {{Google}},
  title        = {{Agent Development Kit (ADK)}: An Open-Source Framework for Building {AI} Agents},
  howpublished = {\url{https://google.github.io/adk-docs/}},
  year         = {2025},
}

@misc{gsd2025,
  author       = {T\^{A}CHES},
  title        = {{Get Shit Done}: A Light-Weight Meta-Prompting, Context Engineering and Spec-Driven Development System for {AI} Coding Agents},
  howpublished = {\url{https://github.com/gsd-build/get-shit-done}},
  year         = {2025},
  note         = {Initial commit December 2025},
}

@misc{anthropic2025prompting,
  author       = {{Anthropic}},
  title        = {Anthropic Prompt Engineering Guide},
  howpublished = {\url{https://docs.anthropic.com/en/docs/build-with-claude/prompt-engineering/overview}},
  year         = {2025},
}

@misc{openai2025prompting,
  author       = {{OpenAI}},
  title        = {Prompt Engineering Best Practices},
  howpublished = {\url{https://platform.openai.com/docs/guides/prompt-engineering}},
  year         = {2025},
}

@misc{google2025prompting,
  author       = {{Google}},
  title        = {Gemini Prompt Engineering Strategies},
  howpublished = {\url{https://ai.google.dev/gemini-api/docs/prompting-strategies}},
  year         = {2025},
}

@inproceedings{opsahlong2024mipro,
  author    = {Opsahl-Ong, Krista and Ryan, Michael J. and Purtell, Josh and Broman, David and Potts, Christopher and Zaharia, Matei and Khattab, Omar},
  title     = {Optimizing Instructions and Demonstrations for Multi-Stage Language Model Programs},
  booktitle = {Proceedings of the 2024 Conference on Empirical Methods in Natural Language Processing (EMNLP)},
  year      = {2024},
  note      = {arXiv preprint arXiv:2406.11695},
}

@inproceedings{yang2024opro,
  author    = {Yang, Chengrun and Wang, Xuezhi and Lu, Yifeng and Liu, Hanxiao and Le, Quoc V. and Zhou, Denny and Chen, Xinyun},
  title     = {Large Language Models as Optimizers},
  booktitle = {International Conference on Learning Representations (ICLR)},
  year      = {2024},
  note      = {arXiv preprint arXiv:2309.03409},
}

@article{fernando2023promptbreeder,
  author    = {Fernando, Chrisantha and Banarse, Dylan and Michalewski, Henryk and Osindero, Simon and Rockt{\"a}schel, Tim},
  title     = {{Promptbreeder}: Self-Referential Self-Improvement via Prompt Evolution},
  journal   = {arXiv preprint arXiv:2309.16797},
  year      = {2023},
}

@inproceedings{pryzant2023apo,
  author    = {Pryzant, Reid and Iter, Dan and Li, Jerry and Lee, Yin Tat and Zhu, Chenguang and Zeng, Michael},
  title     = {Automatic Prompt Optimization with ``Gradient Descent'' and Beam Search},
  booktitle = {Proceedings of the 2023 Conference on Empirical Methods in Natural Language Processing (EMNLP)},
  year      = {2023},
  note      = {arXiv preprint arXiv:2305.03495},
}

@article{yuksekgonul2024textgrad,
  author    = {Yuksekgonul, Mert and Bianchi, Federico and Boen, Joseph and Liu, Sheng and Huang, Zhi and Guestrin, Carlos and Zou, James},
  title     = {{TextGrad}: Automatic ``Differentiation'' via Text},
  journal   = {arXiv preprint arXiv:2406.07496},
  year      = {2024},
}

@article{ouyang2025kernelbench,
  author    = {Ouyang, Anne and Guo, Simon and Arora, Simran and Zhang, Alex L. and Hu, William and R{\'e}, Christopher and Mirhoseini, Azalia},
  title     = {{KernelBench}: Can {LLMs} Write Efficient {GPU} Kernels?},
  journal   = {arXiv preprint arXiv:2502.10517},
  year      = {2025},
}

@misc{anthropic2026remotecontrol,
  author       = {{Anthropic}},
  title        = {{Claude Code Remote Control}: Continue Local Sessions from Any Device},
  howpublished = {\url{https://docs.anthropic.com/en/docs/claude-code/remote-control}},
  year         = {2026},
}

@misc{github2026copilotcloud,
  author       = {{GitHub}},
  title        = {Using {Copilot} Cloud Agent on {GitHub}},
  howpublished = {\url{https://docs.github.com/en/copilot/how-tos/use-copilot-agents/cloud-agent/use-cloud-agent-on-github}},
  year         = {2026},
}

@misc{cursor2026cloudagents,
  author       = {{Cursor}},
  title        = {{Cursor} Cloud Agents},
  howpublished = {\url{https://cursor.com/docs/cloud-agent}},
  year         = {2026},
}

@misc{openai2026codexchangelog,
  author       = {{OpenAI}},
  title        = {{Codex} Changelog},
  howpublished = {\url{https://developers.openai.com/codex/changelog/}},
  year         = {2026},
}

@misc{aider2026watch,
  author       = {Gauthier, Paul},
  title        = {{Aider} in Your {IDE}: {AI} Comments and File Watching},
  howpublished = {\url{https://aider.chat/docs/usage/watch.html}},
  year         = {2026},
}

@misc{langchain2026hitl,
  author       = {{LangChain}},
  title        = {Human-in-the-Loop Middleware},
  howpublished = {\url{https://docs.langchain.com/oss/python/langchain/human-in-the-loop}},
  year         = {2026},
}

@misc{microsoft2025semantickernelplanning,
  author       = {{Microsoft}},
  title        = {Planning in {Semantic Kernel}},
  howpublished = {\url{https://learn.microsoft.com/en-us/semantic-kernel/concepts/planning}},
  year         = {2025},
}
}

\end{document}